\documentclass{article}
\usepackage[abbr]{harvard}
\usepackage[T1]{fontenc}            % for german characters
\usepackage{amsmath,amssymb}
\usepackage[german,english]{babel}  % for german hyphenation, quotes, etc. last language is default

\makeatletter
\author{Enric P\'erez\\
{\footnotesize Departament de F\'isica Fonamental, Universitat de Barcelona}\\[-0.1cm]
{\footnotesize c. Mart\'i i Franqu\`es 1, 08028 Barcelona, Spain}\\[-0.1cm]
{\footnotesize enperez@ub.edu}\\
\and Tilman Sauer\\
{\footnotesize Einstein Papers Projects, Caltech 20-7}\\[-0.1cm]
{\footnotesize Pasadena, CA91125, USA}\\[-0.1cm]
{\footnotesize tilman@caltech.edu}
}%author
\makeatother

\date{Version of \today}

\title{Einstein's quantum theory of the monatomic ideal gas:\\
	non-statistical arguments for a new statistics}

\begin{document}

\maketitle

\begin{abstract}
In this article, we analyze the third of three papers, in which Einstein presented his quantum theory of the ideal gas of 1924--1925. Although it failed to attract the attention of Einstein's contemporaries and although also today very few commentators refer to it, we argue for its significance in the context of Einstein's quantum researches. It contains an attempt to extend and exhaust the characterization of the monatomic ideal gas without appealing to combinatorics. Its ambiguities illustrate Einstein's confusion with his initial success in extending Bose's results and in realizing the consequences of what later became to be called Bose-Einstein statistics. We discuss Einstein's motivation for writing a non-combinatorial paper, partly in response to criticism by his friend Ehrenfest, and we paraphrase its content. Its arguments are based on Einstein's belief in the complete analogy between the thermodynamics of light quanta and of material particles and invoke considerations of adiabatic transformations as well as of dimensional analysis. These techniques were well-known to Einstein from earlier work on Wien's displacement law, Planck's radiation theory, and the specific heat of solids. We also investigate the possible role of Ehrenfest in the gestation of the theory.\end{abstract}
\newpage

\tableofcontents

\section{Introduction}

It has been said that Albert Einstein's quantum theory of the monatomic ideal gas, the 
conceptual innovation of Bose-Einstein statistics in the quantum physics of material particles, 
was his last ``positive contribution'' to statistical physics.%
\footnote{\cite[p.~175]{BornM1949Theories}. In a similar vein, Pais takes the work on the 
quantum ideal gas to be the last valid achievement in Einstein's intellectual career, when he 
suggests that his fame was ``based exclusively on what he did before 1925,'' in the infamous 
dictum about Einstein's later biography that ``his fame would be undiminished, if not enhanced, 
had he gone fishing instead.'' \cite[p.~43]{PaisA1994Einstein}.}
It was presented in three papers published in 1924 and 1925.%
\footnote{\cite{EinsteinA1924Quantentheorie,EinsteinA1925Quantentheorie1,EinsteinA1925Quantentheorie2}.}  In these papers, Einstein made an important step in the quantization of the ideal 
gas, i.e.\ of a system of free, massive particles confined in a volume.

The historical connections of Einstein's theory with earlier work by Satyendra Nath Bose, on the 
one hand, and with Erwin Schr\"odinger's wave mechanics, on the other hand, have already been 
widely discussed in the literature.  

Most historical commentary focuses on Einstein's first two papers, which indeed contain the 
most significant conclusions of the theory: a new distribution law for the energy, a new way of 
counting microstates, an analysis of fluctuations, and the prediction of what came to be known 
as the Bose-Einstein condensation phenomenon. The third paper, in contrast, has rarely been 
mentioned, and we have not found any work that would analyze it in some detail. 
Max Jammer\footnote{\cite{JammerM1966Development}.},
Friedrich Hund\footnote{\cite{HundF1975Geschichte}.},
Abraham Pais\footnote{\cite{PaisA1982Subtle}.},
Jagdish Mehra and Hans Rechenberg,\footnote{\cite{MehraJEtAl1982Development1,MehraJEtAl1984Development4}.}
and Olivier Darrigol\footnote{\cite{DarrigolO1991Statistics}.},
for example, cite the paper but do not comment on it,  i.e., they refer to the list of all three 
publications, but limit their comments to the results of the first two papers only.\footnote{Works that contain discussion of Einstein's first two notes but fail to mention the third paper include \cite{EzawaH1979Contribution}.}
Martin Klein, in a reference article on Einstein and the wave-particle duality, does not even cite the third paper.%
\footnote{\cite{KleinM1964Einstein}. He did cite the third paper in an interesting article on
Ehrenfest's contributions to the development of quantum statistics \cite{KleinM1959Contributions,KleinM1959Contributions2}.}

Agostino Desalvo, in a long paper, in which he analyzed different attempts of calculating the chemical 
constant and their relationship to the birth of quantum statistics, discussed Einstein's third 
paper, albeit only briefly. In fact, his comments suggest that the paper deserves closer attention:

\begin{quote}
This paper usually receives less consideration than the former two. However, if one recalls the 
key role of thermodynamics in Einstein's thought and the discussion of thermodynamics 
requirements imposed on the theory of gas degeneracy (...) this paper appears to be a 
\emph{necessary} complement to the other two.%
\footnote{\cite[p.~526]{DesalvoA1992Constant}. His emphasis.}
\end{quote}

Einstein followed an approach in this paper that was not based only on statistical considerations and 
that was closer to thermodynamics.
He tried to find general conditions that any theory of the 
ideal gas would have to satisfy, mainly by establishing and exploiting analogies with radiation, 
where the displacement law at least provided some hints as to what the radiation law should 
look like.%
\footnote{In the bibliography compiled by Margaret Shields for the book \emph{Albert Einstein: 
Philosopher-Scientist} this paper is described as follows: ``A general condition is deduced which 
must be satisfied by every theory of a perfect gas.'' \cite[p.~716]{SchilppPA1949Einstein}. The 
phrase is almost a literal quote from Einstein's paper. See footnote \ref{note:lastpara} .}

Paul Hanle, in a general survey of Schr\"odinger's research on statistics of ideal gases prior to 
the formulation of wave mechanics,%
\footnote{\cite{HanlePA1977Coming}.}
represents another exception.  To be sure, his comments 
are not any more explicit than Desalvo's. He suggested one should understand Einstein's third 
paper as a response to Paul Ehrenfest's objections against the reality of the condensation 
phenomenon. But he also suggested that Ehrenfest was not the only addressee and that 
Einstein took ``Ehrenfest's criticism as symptomatic of scepticism towards the theory among his 
colleagues.''%
\footnote{Ibid., p.~176--177.}

In summary, Einstein's third paper has received very little attention from historians. Neither did it 
receive a lot of attention at the time of its publication. We hardly have found references to it by 
contemporaries, and references to the paper are scarce even by Einstein himself.

From a historical point of view, the fact that Einstein wrote a non-combina\-torial paper after 
expounding his new theory of the quantum ideal gas in two prior articles points to a deeper 
conceptual problem. There are indications that Einstein himself may not have realized the full 
implications of the new way of counting, despite his earlier work on black-body radiation. For 
example, Daniela Monaldi has argued in a note on the prehistory of indistinguishable 
particles that ``neither Bose nor Einstein showed any awareness that they were inaugurating the 
statistics of indistinguishable particles.''%
\footnote{\cite[p.~8]{MonaldiD2009Note}.}
Such observations raise a methodological problem. Indeed, careful reading of Bose's paper as 
well as of Einstein's first two notes do not, it seems to us, allow a modern reader to decide 
whether Einstein or Bose were fully aware, at the time, of the conceptual implications of their 
new way of counting. We do have, however, parts of an epistolary exchange between Einstein 
and the Viennese physicist Otto Halpern.\footnote{See footnote \ref{note:Halpernletter}.} The correspondence was initiated by Halpern in 
response to Einstein's note, and in it we find a very explicit discussion of the new combinatorics, 
both by Halpern and by Einstein.
While it therefore seems that Einstein became aware of the 
implications of the new conceptual implications of Bose-Einstein statistics, at least, in the period 
between the publication of the first and the second paper, we also have explicit criticism by his 
colleague Paul Ehrenfest, which points to the fact that the new way of statistics was rejected just 
because of these novel implications.

As we will elaborate in this article, the implications of indistinguishability were discussed at 
the time under the label of ``loss of statistical independence.''

For a historical reconstruction of the emergence of one of the core conceptual innovations of 
quantum theory, it is therefore of interest to take a close look at Einstein's third paper on the 
quantum ideal gas, precisely because it set out to justify this new theory without making use of 
the new combinatorics.

Our interest in the non-statistical paper on the quantum ideal gas arose initially from our interest 
in Paul Ehrenfest's adiabatic hypothesis and, more generally, in his work. In the third paper, 
Einstein used an adiabatic transformation as a part of a process designed to provide an 
argument to support his new theory of the quantum ideal gas. Indeed, as we will show, a 
detailed analysis of the paper suggests other interesting relations to Ehrenfest's research. It is 
well known and has been observed before%
\footnote{\cite[p.~430]{PaisA1982Subtle}.}  
that Einstein mentioned his good friend in the second paper, but only in relation to the question 
of loss of statistical independence of the particles.%
\footnote{See note \ref{note:PEmention}.}
The discovery of a manuscript of that second paper in the professional library of 
Ehrenfest in Leiden%
\footnote{\cite{HuijnenPEtAl2007Road}.}
further kindled our interest in what appears to have been a debate between the two physicists in  
the---more or less---six months of gestation that preceded this third contribution by Einstein on 
the quantum ideal gas. 

In view of all this, our intention is to analyze the content of the third paper without any further 
analysis of the pair that preceded it, since they have already been studied in detail.%
\footnote{See, for instance, \cite{NavarroL2009Einstein} and references therein.}
We will try to account for its gestation period, in particular as regards the role that Ehrenfest 
would have taken in it and also compare it with  previous and later reflections by Einstein 
himself. Finally, we will formulate some conjectures as to why this paper met cold reception 
despite its historical and systematic interest.

In the title of this essay, we refer to the third paper as containing ``non-statistical arguments.'' More 
accurately, it should state ``non-combinatorial arguments.'' 
In a certain sense, as we will see, 
it does contain some statistical results, insofar as it deals with the distribution function of the 
kinetic energy among the molecules. However, that function is not analyzed starting from the 
microscopic constituents of the system, but deduced from its macroscopic properties. What 
Einstein really omitted completely in this paper is any argument of combinatorics. We have kept 
the word `statistical' in the title because it was the consideration of the kind of dependence 
among molecules which Einstein tried to avoid. It was a non-statistical paper in the sense that 
the way the microstates had to be counted was not discussed.

In Box 1 (p.~\pageref{table:chrono}) we have gathered some of the relevant dates for what follows and to which we 
will refer throughout the paper.

\begin{table}
\label{table:chrono}
\begin{tabular}{|p{3cm}|p{8.2cm}|}
\hline 
4 June 1924	& Bose writes to Einstein\\
c. 2 July 1924	& Bose's paper (translated by Einstein) received by \emph{Zeitschift f\"ur Physik}\\
10 July 1924	& Einstein's first paper on QTMIG presented to the Prussian Academy (PA)\\
20 September 1924& Einstein's first paper on QTMIG published \cite{EinsteinA1924Quantentheorie}\\
December 1924	& Einstein's second paper on QTMIG signed\\
& Bose's paper published \cite{BoseSN1924Gesetz}\\
8 January 1925	& Einstein's second paper on QTMIG presented to PA\\
29 January 1925	& Einstein's third paper on QTMIG presented to PA\\
9 February 1925 & 	Einstein's second paper on QTMIG published \cite{EinsteinA1925Quantentheorie1}\\
5 March 1925	& Einstein's third paper on QTMIG published \cite{EinsteinA1925Quantentheorie1}\\
\hline
\end{tabular}
\caption{Chronology of the presentation and publication of Einstein's quantum theory of
the monoatomic ideal gas (QTMIG) and some related facts}
\end{table}

\section{Einstein's quantum theory of the monatomic ideal gas}

Some time in June 1924, Einstein received a letter from a Bengali physicist, Satyendra Nath 
Bose, who asked him politely to translate---if he believed it was worth it---and forward for 
publication a paper on the hypothesis of light quanta, which he had attached.%
\footnote{See \cite{BlanpiedWA1972Bose} and \cite{WaliKC2006Man}. The editors of 
	\emph{Philosophical Magazine} had earlier rejected Bose's manuscript.}
Einstein complied 
and translated and sent to \emph{Zeitschrift f\"ur Physik} Bose's subsequently famous paper.%
\footnote{\cite{BoseSN1924Gesetz}.}
To the published paper, he added the following commentary:

\begin{quote}
In my opinion Bose's derivation of the Planck formula signifies an important advance. The 
method used also yields the quantum theory of the ideal gas, as I will work out in detail 
elsewhere.%
\footnote{\selectlanguage{german}
	``Bose's Ableitung der Planckschen Formel bedeutet nach meiner Meinung einen wichtigen
	Fortschritt. Die hier benutzte Methode liefert auch die Quantentheorie des idealen
	Gases, wie ich an anderer Stelle ausf\"uhren will.'' \cite[p.~181]{BoseSN1924Gesetz}.
	An English translation of \cite{BoseSN1924Gesetz} can be found in \cite[p.~1056]{TheimerOetAl1976Beginning}.}
\end{quote}

In Bose's paper we find, for the first time, a derivation of the factor
\begin{equation}
\frac{8\pi\nu^2}{c^3}Vd\nu,
\label{eq:prefactor}
\end{equation}
starting from the quantization of energy ($c$ is the speed of light in vacuum, $V$ the volume). This 
expression gives the number of cells corresponding to frequencies between $\nu$ and $\nu+d\nu$ or, in wave-theoretical terms, the number of modes with frequency in that same range. With the average energy of a resonator of frequency $\nu$ (or of a normal mode) it constitutes Planck's blackbody radiation law for the energy density $r$:
\begin{equation}
r(\nu,T)d\nu=\frac{8\pi\nu^2}{c^3}\frac{h\nu}{e^{\frac{h\nu}{kT}}-1}d\nu
\end{equation}
($k$ is Boltzmann's constant and $T$ the temperature). 
While several different ways had been found to derive the average energy of a resonator based on Planck's quantum hypothesis, the prefactor had previously been derived only classically, without 
invoking the concept of quantization. Thus, in his 1916 paper Einstein remarked about the prefactor:
\begin{quote}
In order to obtain the numerical value of constant $\alpha$ [defined earlier as $\rho = \alpha\nu^31/(\exp(-h\nu/kT)-1)$] one would have to have an exact theory of electrodynamic and mechanical processes. For the time being we must use Rayleigh's limiting case of high temperatures, for which the classical theory applies in the limit.%
\footnote{\selectlanguage{german}
	``Um den numerischen Wert der Konstante $\alpha$ zu ermitteln, m\"u{\ss}te man eine exakte Theorie der elektrodynamischen und mechanischen Vorg\"ange haben; man bleibt hier vor\"aufig auf die Behandlung des Rayleigh'schen Grenzfalles hoher Temperaturen angewiesen, f\"ur welchen die klassische Theorie in der Grenze gilt.'' \cite[p.~53]{Einstein1916n}. Note that Einstein's notation in this quote is inconsistent with the one that we use throughout. In this article, we use $r$ to denote the distribution function for radiation and $\rho$ for material gases.}
\end{quote}
In order to obtain this factor, Bose divided the six-dimensional phase space of a light 
quantum into cells of (hyper)volume $h^3$. He then calculated the probability of a macroscopic 
state, taking as a microstate only the number of quanta that were contained in each cell, 
disregarding any information as to which individual quanta were contained in which cell. With 
that move, and by just applying the orthodox methods inherited from Ludwig Boltzmann, he was 
able to derive Planck's radiation law. 

We will not give any more details about Bose's bold idea and his paper because it is discussed 
at length elsewhere.%
\footnote{See, e.g., \cite{KleinM1964Einstein}, \cite{BergiaS1983Who}.}
Nowadays, Bose's discovery is mostly presented as a striking example of 
serendipity, since it seems that its author was not fully aware of the significance of the step he 
was taking.%
\footnote{See, e.g., \cite{DelbrueckM1980Was}, \cite[pp.~424--428]{PaisA1982Subtle},
	\cite{BergiaS1983Who}.} 

Einstein was aware of the significance, as is evident from the swiftness with which he 
translated and submitted Bose's paper, and from the footnote that he attached to it and 
that we have quoted above. In fact, before receiving Bose's manuscript he had recently returned
himself to an 
investigation of the theory of light quanta. On 24 April 1924, Einstein gave a presentation in the 
plenary session of the Prussian Academy of Sciences ``about the present state of radiation 
problem.''%
\footnote{\emph{Sitzungsberichte Preu{\ss}ische Akademie der Wissenschaften,
	Physikalisch-mathematische Klasse}, 1924, 179. The abstract of the \emph{Sitzungsberichte}
	indicates the content of Einstein's talk: ``Statistical properties of radiation. Discussion
	of Bothe's theory of multiple quanta and of attempts by the author to solve the quantum
	problem by means of overdetermined systems of equations'' (``Statistische Eigenschaften
	der Strahlung. Betrachtung \"uber Bothes Theorie der mehrfachen Quanten und \"uber 
	Bem\"uhungen des Verfassers, das Quantenproblem durch \"uberbestimmte Gleichungssystem zu
	l\"osen.'') The reference to Einstein's own work presumably is to \cite{EinsteinA1923Bietet}.
	The abstract in the \emph{Sitzungsberichte} is preceded with a little star (see also
	the manuscript for the abstract, AEA 05-187, available at www.alberteinstein.info), which
	indicates that the report was not intended for publication, at least not by the Academy.
	This implicit use of a star for titles and abstracts of presentations to the Academy listed
	in its \emph{Sitzungsberichte} had been common since 1902. In earlier issues of the
	\emph{Sitzungsberichte}, the meaning of the star had been made explicit at the bottom of
	the page, but during the year 1902, the explicit footnote attached to the star began
	to be dropped. }
Only a few weeks before receiving Bose's manuscript, he wrote to a friend:
\begin{quote}
As regards scientific work, I am pondering almost exclusively the quantum problem and I now 
believe to be really on the right track, if it is certain. The best I had achieved in these matters in 
recent times was the work of 1917 in the \emph{physikal.\ Zeitschrift}. 
My new efforts aim at unification 
of quanta and Maxwell's field. Among the experimental results of recent years, it is only the 
experiments by Stern and Gerlach and the experiment by Compton (scattering of R\"ontgen 
radiation together with a change of frequency) that are of any significance. The first one proves 
the independent existence of the quantum states, the second one proves the reality of the 
momentum of light quanta.%
\footnote{\selectlanguage{german}
	``Wissenschaftlich h\"ange ich fast ununterbrochen dem Quantenproblem nach und
	glaube wirklich auf der richtigen Spur zu sein---wenns gewiss ist. Das Beste was
	mir da in sp\"aterer Zeit gelungen ist, war die Arbeit von 1917 in der
	physikal.\ Zeitschrift. Meine neuen Bestrebungen gehen auf Vereinigung von Quanten
	und Maxwell'schen Felde. Von den experimentellen Ergebnissen der letzten Jahre
	sind eigentlich nur die Experimente von Stern und Gerlach sowie das Exp.\ von Compton
	(Zerstreuung der R\"ontgenstrahlung mit Frequenz\"anderung) von Bedeutung, deren erstes
	die Allein-Existenz der Quantenzust\"ande, deren zweites die Realit\"at des Impulses
	der Lichtquanten beweist.'' Albert Einstein to Michele Besso, 24 May 1924. In
	\cite[p.~202]{SpezialiP1972Correspondance}, (French paperback edition, p.120)}
\end{quote}

Therefore, Bose's manuscript was timely: After the experimental successes by Arthur Compton 
and Peter Debye, which seemed to confirm that light quanta have momentum as well as 
energy;%
\footnote{See, e.g., \cite[pp.~512-532]{MehraJEtAl1982Development1} for a historical discussion.}
after the spectacular discovery of Otto Stern and Walther Gerlach, for many 
physicists---Einstein among them---the most striking and convincing demonstration of 
quantization;%
\footnote{See ibid., pp.~422-445 for a historical discussion.}
and shortly after Einstein's return to his own research on light quanta. 
Probably for this reason it took him so little time to prepare a presentation in which he applied 
Bose's method to an ideal gas.%
\footnote{\cite{EinsteinA1924Quantentheorie}.}
He presented it at the Prussian Academy on 10 July, only a 
month after Bose had signed his letter.%
\footnote{In a comparable situation, Einstein surprised his colleague David Hilbert with a swift 
	calculation of the anomalous advance of Mercury's perihelion after giving up his \emph{Entwurf}-equations and reverting to generally covariant field equations, see Hilbert to Einstein, 19 November 1915 \cite[Doc.~149]{CPAE08}: ``If I could do the calculations as rapidly as you, the electron would have to
surrender and the hydrogen atom would have to produce a letter from home
excusing it from not radiating.'' The background for Einstein's achievement was, of course, that he had done before detailed calculations of the perihelion problem in the context of the \emph{Entwurf}-theory, which he could readily assimilate to the case of the new field equations, see \cite{EarmanJEtAl1993Explanation} for a detailed discussion.}

In this paper we find the density of states of (kinetic) energy $E$ for a molecule of mass $m$ of an 
ideal gas:
\begin{equation}
2\pi \frac{V}{h^3}(2m)^{\frac{3}{2}}E^{\frac{1}{2}}dE,
\label{eq:densityofstates}
\end{equation}
which is the analogue of (1): It gives the number of phase cells of a single molecule 
corresponding to energies between $E$ and $E + dE$. Following Bose's derivation, Einstein 
maximized the probability of a certain distribution of molecules in phase space, which he had 
previously divided into cells of volume $h^3$. He also took into account only how many molecules 
were in each cell, not which, and introduced the constraint of the total number of particles, a 
condition that is not invoked in the case of light quanta. He obtained the average occupation 
number of a state with energy $E$, and also  the equation of state of the ideal gas:
\begin{equation}
p=\frac{2}{3}\frac{\overline{E}}{V}
\label{eq:penerg}
\end{equation}
($p$ is the pressure and $\overline{E}$  the mean energy of the gas).
He commented on this result with the 
remark: ``We obtain the notable result that the relation between kinetic energy and pressure is 
exactly the same as in the classical theory, where it is derived from the virial theorem.''%
\footnote{\selectlanguage{german}
	``Es ergibt sich also das merkw\"urdige Resultat, dass die Beziehung zwischen der 
	kinetischen Energie und dem Druck genau gleich herauskommt wie in der klassischen Theorie,
	wo sie aus dem Virialsatz abgeleitet wird.'' \cite [p.~264]{EinsteinA1924Quantentheorie}.}
We will see below that in the third instalment of his theory Einstein tried to take advantage of this 
coincidence.

In this seminal paper, Einstein also showed how classical results can be obtained by an 
expansion of expressions corresponding to the new theory in power series of
a parameter $\lambda$, defined as
\begin{equation}
\lambda \equiv \frac{h^3}{\pi^{\frac{3}{2}}(2\pi m\kappa T)^{\frac{3}{2}}}\frac{N}{V},
\end{equation}
and by keeping only the first term ($\lambda \ll 1$). 
He wrote some expressions that allowed him to see 
the differences between both theories to that order of approximation. For instance, for the 
average energy of the system he found:
\footnote{We correct the wrong numerical factor $0.0318$ that appears in the paper. Einstein himself corrected this mistake in the last paragraph of his next paper, without, however, pointing it out, see \cite[p.~13]{EinsteinA1925Quantentheorie1}. Desalvo and Navarro have already noted this omission \cite[p.~524]{DesalvoA1992Constant}, \cite[pp.~200--201]{NavarroL2009Einstein}.} 
\begin{equation}
\frac{\overline{E}}{N} =
\frac{3}{2}\kappa T \left[1-0.1768 h^3\frac{N}{V} (2\pi m\kappa T)^{-\frac{3}{2}}\right]
\end{equation}

Einstein pointed out that contrary to what happens in the ordinary theory, the new expression 
for the entropy of the gas is perfectly compatible with Nernst's principle, in the sense that the 
entropy vanishes at zero temperature. In fact, in Einstein's theory, at zero temperature, all 
molecules are in the same cell, leaving only one microstate possible.

At the end of the paper we find an interesting comment on a question that had been and still 
was widely discussed by his predecessors in the study of the quantum ideal gas: The Gibbs' 
paradox.%
\footnote{Einstein only refers to ``a paradox'' (``ein Paradoxon'') and does not identify it as Gibbs' paradox. We have no evidence for assuring that Einstein knew Gibbs' paradox. However, Ehrenfest did, see footnote~{\ref{note:Ehrenfestdiaries}}. Moreover, Schr\"odinger had recently published a paper in the \textit{Zeitschrift f\"ur Physik} entitled ``Isotopie und Gibbsches Paradoxon'' \cite{SchroedingerE1921Isotopie}. It is more than likely that Einstein had heard about it before writing this paragraph.}
In Einstein's theory the entropy of the gas is extensive and, like the classical entropy, 
additive with respect to different components. If the mixture of two different gases implies an 
increase of entropy, the \textit{mixture} of the same gas (at equal density), on the other hand, does 
not. According to Einstein, this prevents one from imagining a continuous variation of the 
differences between gases.

In the second instalment Einstein proposed a solution to this question.%
\footnote{\cite{EinsteinA1925Quantentheorie1}.}
The second paper was 
signed in December and read at the Academy's meeting of 8 January 1925. Since the 
presentation of the previous paper, Einstein had had plenty of time pondering and discussing 
the subject with his colleagues. The second paper presents further detailed analysis of the 
consequences implied by the theory expounded in the first paper. Einstein emphasized this fact 
by numbering both equations and paragraphs in consecutive order with the first one (the 
second paper begins with the sixth paragraph and with the 24th equation).%
\footnote{``For convenience, I write the following formally as a continuation of the paper cited.''
	(``Der Bequemlichkeit halber schreibe ich das Folgende formal als Fortsetzung der 
	 zitierten Abhandlung.'') \cite[p.~3]{EinsteinA1925Quantentheorie1}.}
The most famous 
results of Einstein's theory are contained in this paper. In this paper indeed, Einstein took the 
theory considerably further than Bose had done. 

First, Einstein discussed an unusual consequence: the condensation at low temperatures or, in 
other words, the saturated gas. Einstein considered, for the first time, the case of a gas in 
which, below a certain critical temperature (that depends on $N$ and $V$), the number of particles in 
excited states is limited. In the next section, he discussed the loss of statistical independence of 
the molecules in a famous passage where Ehrenfest's name appears:

\begin{quote}
Mr. Ehrenfest and other colleagues have raised the criticism that in Bose's theory of radiation 
and in my analogous theory of ideal gases the quanta or molecules are not treated as 
statistically independent entities without explicit mentioning of this feature in our respective 
papers. This is entirely correct.
\end{quote}
And the passage continues: 

\begin{quote}
If the quanta are treated as statistically independent regarding their localization, one obtains 
Wien's law of radiation; if one treats the gas molecules in an analogous way, one arrives at the 
classical equation of state, even if one proceeds in exactly the same way as Bose and I have 
done.%
\footnote{\selectlanguage{german}
	``Von {\scshape Ehrenfest} und anderen Kollegen ist an {\scshape Boses} Theorie der
	Strahlung und an meiner analogen der idealen Gase ger\"ugt worden, da{\ss} in diesen
	Theorien die Quanten bzw. Molek\"ule nicht als voneinander statistisch unabh\"angige
	Gebilde behandelt werden, ohne da{\ss} in unseren Abhandlungen auf diesen Umstand
	besonders hingewiesen worden sei. Dies ist v\"ollig richtig. Wenn man die Quanten als
	voneinander statistisch unabh\"angig in ihrer Lokalisierung behandelt, gelangt man zum
	{\scshape Wien}schen Strahlungsgesetz; wenn man die Gasmolek\"ule analog behandelt,
	gelangt man zur klassischen Zustandsgleichung der idealen Gase, auch wenn man im \"ubrigen
	genau so vorgeht, wie {\scshape Bose} und ich es getan haben.'' Ibid., 5. 
	\label{note:PEmention}}
\end{quote}

Then, Einstein elucidated this issue analytically, but he left in the dark what kind of dependence 
it is that affects the behaviour of molecules in the new statistics. He pointed out something that 
he had already suggested in his previous paper: In classical theory the entropy expression 
forces one to choose between two different conditions to be fulfilled, that is, Nernst's principle or 
the extensivity of entropy. In the new theory, the two conditions are satisfied at the same time. 
Einstein considered this fact a strong support of the deep analogy between radiation and gas on 
which his theory was founded:

\begin{quote}
For these reasons I believe that one has to prefer the conception a) (i.e., Bose's statistical 
approach) even if this preference over others cannot be justified apriori. This result in itself 
lends support for the belief in the deep essential similarity between radiation and gas in that the 
same statistical conception that leads to Planck's formula produces the agreement between gas 
theory and Nernst's theorem when applied to ideal gases.%  
\footnote{\selectlanguage{german}
	``Aus diesen Gr\"unden glaube ich, dass der Berechnungsweise a) (d.h. Boses
	statistischem Ansatz) der Vorzug gegeben werden muss, wenn sich die Bevorzugung dieser
	Berechnungsweise anderen gegen\"uber auch nicht a priori erweisen l\"asst. Dies Ergebnis
	bildet seinerseits eine St\"utze f\"ur die Auffassung von der tiefen Wesensverwandtschaft
	zwischen Strahlung und Gas, indem dieselbe statistische Betrachtungsweise, welche zur
	Planckschen Formel f\"uhrt, in ihrer Anwendung auf ideale Gase die \"Ubereinstimmung der
	Gastheorie mit dem Nernstschen Theorem herstellt.''  Ibid., 7}
\end{quote}

Also in this paper, we find the first appeal by Einstein to a certain duality in terms of the thesis by 
Louis de Broglie. After analyzing the energy fluctuations of an ideal gas, he described the ideas 
of the French physicist aimed at overcoming the opposition between waves and particles. The 
great impact this reference by Einstein to de Broglie's work had on the research of Schr\"odinger 
has been noted on many occasions, as Schr\"odinger never failed to recognize it.%
\footnote{See \cite{KleinM1964Einstein}, \cite{SchroedingerE1926Uebergang}.
	Also, Schr\"odinger to Einstein, 23 April 1926. 
	English translation in \cite[p.~26]{PrzibramK1967Letters}.}
Appealing to 
the wave field that would accompany each particle, Einstein proposed to solve the paradox with 
which he had closed the previous paper: The interference will only take place in gases 
composed of molecules of equal mass.

Finally, Einstein suggested two effects of his theory that were possibly accessible to 
experimental verification. The first one is a decrease in viscosity. The undulatory behaviour of 
the molecules should lead to diffraction effects that might provoke, in gases of low-mass 
elements such as helium or molecular hydrogen, a dramatic decrease in the friction coefficient 
of the gas. But after calculating the size of the required apertures, Einstein discards standard 
diffraction experiments for this effect. Second, he proposed to use the statistics of a saturated 
gas to account for the problem why the electronic contribution to the specific heat of metals is so 
low. However, in this case, Einstein admits that the difficulties in applying this idea are so big 
that it can hardly be considered a proof of his theory. 

We regard Einstein's second paper on the quanta a milestone in the history of quantum 
physics, not only because of the unusual amount of new results it contains but also because in a 
certain sense it closed the circle that was initiated by Einstein himself twenty years earlier with 
his heuristic hypothesis of light quanta. He was a pioneer in emphasizing the dual nature of 
radiation in 1909. In 1925, with a completely analogous procedure, he in turn demonstrated the 
validity of his proposal for the ideal gas.

In short, Einstein developed the analogy between gas and radiation, knowing that despite the 
evidence he could adduce to support the theory, it was unsure whether his theory was the true 
theory. In his own words:
 
\begin{quote}
The interest in this theory derives from the fact that it is based on the hypothesis of an extended 
formal similarity between radiation and gas. According to this theory, the degenerate gas differs 
from the gas of mechanical statistics in an analogous way as the radiation according to Planck's 
law differs from the radiation according to Wien's law. If one takes Bose's derivation of Planck's 
radiation formula seriously, then one cannot ignore this theory of the ideal gas either; because if 
it is justified to conceive of the radiation as a gas of quanta, then the analogy between a gas of 
quanta and a gas of molecules must be a complete one.%
\footnote{\selectlanguage{german}
	``Das Interesse dieser Theorie liegt darin, da{\ss} sie auf die Hypothese einer
	weitgehenden formalen Verwandschaft zwischen Strahlung und Gas gegr\"undet ist. Nach dieser
	Theorie weicht das entartete Gas von dem Gas der mechanischen Statistik in analoger Weise
	ab wie die Strahlung gem\"a{\ss} dem {\scshape Planck}schen Gesetze von der Strahlung
	gem\"a{\ss} dem {\scshape Wien}schen Gesetze. Wenn die {\scshape Bose}sche Ableitung der
	{\scshape Planck}schen Strahlungsformel ernst genommen wird, so wird man auch an dieser
	Theorie des idealen Gases nicht vorbeigehen d\"urfen; denn wenn es gerechtfertig ist, die
	Strahlung als Quantengas aufzufassen, so mu{\ss} die Analogie zwischen Quantengas und
	Molek\"ulgas eine vollst\"andige sein.'' \cite[p.~3]{EinsteinA1925Quantentheorie1}.}
\end{quote}

We finish this brief summary with this quote in order to emphasize the continuity of Einstein's 
strategy. In the third paper, Einstein insisted on this analogy in order to obtain new arguments 
for the validity of the theory, but in this case, as he wrote to Ehrenfest, arguments that were 
independent from the ``incrimininated statistics.''%
\footnote{Einstein to Ehrenfest, 8 January 1925 (AEA 10-097).}

\section{Ehrenfest's role in the prehistory of the third paper}

We will argue that Einstein's third paper is implicitly a response to Ehrenfest's scepticism 
toward Einstein's new theory. When did Ehrenfest learn about Einstein's new theory? Einstein 
first communicated to his friend the discovery in a letter:

\begin{quote}
The Indian Bose gave a beautiful derivation of Planck's law including its constant on the basis of 
the lose light quanta. Derivation elegant, but essence remains obscure. I applied his theory to 
the ideal gas. Rigorous theory of `degeneracy.' No zero point energy and above no energy 
defect. The Lord knows whether it's like this.\footnote{\selectlanguage{german}
	``Der Inder Bose hat eine sch\"one Ableitung des Planckschen Gesetzes samt Konstante
	auf Grund der losen Lichtquanten gegeben. Ableitung elegant, aber Wesen bleibt dunkel.
	Ich habe seine Theorie auf ideales Gas angewendet. Strenge Theorie der
	\glq Entartung.\grq~Keine Nullpunktsenergie und oben kein Energiedefekt. Gott weiss,
	ob es so ist.'' Einstein to Ehrenfest, 12 July 1924 (AEA 10-089). French translation
	published in \cite[p.~166]{OCAE01}.}
\end{quote}

Einstein presented the first instalment on July 10, and this letter was signed on the 12th. At the 
end of month, the two friends could have discussed the matter in person, since Ehrenfest 
stopped over in Berlin for some days in his voyage towards Petersburg. However, during those days 
Einstein was not in Berlin.%
\footnote{Einstein was in Switzerland on 29 July (AEA 143-159), returning to Berlin around 18--20
	August (AEA 120-908) after also visiting Lautrach (AEA 120-907).}
Ehrenfest took part in the fourth congress of the Russian Society of 
Physics (this was his first visit after the Revolution and after he moved to Leiden in 1912).%
\footnote{See \cite[p.~88]{FrenkelVI1971Ehrenfest}.}
In Petersburg, Ehrenfest did not present his friend's new theory, but probably he talked about it 
with some interested physicists, as Joff\'e or Krutkow. On the 18th of September, in his 
presentation on the ``theory of quanta,'' he referred to the struggle that was taking part ``in the 
heart of every physicist'' between corpuscular and undulatory theories.\footnote{\cite{EhrenfestP1924Teoriia}. The text we quote appears in \cite[p.~244] {HallK2008Schooling}.
We wish to thank Karl Hall for sending us the text of the r\'esum\'e of Ehrenfest's talk.}  Joff\'e himself spoke the same day about the ``light atoms.'' 

There is also no evidence that Ehrenfest met Einstein on his return trip to Leiden at the end of 
September or beginning of October%
\footnote{Again, Einstein was out of town, arriving in Vienna on 22 September (AEA 92-097) and
	travelled to Leiden via Innsbruck (AEA 143-163) and Lucerne (AEA 84-567). Nevertheless,
	Ehrenfest may have visited Einstein's wife and step-daughters on his stopover
	in Berlin, see, e.g., (AEA 143-168).}
(after having spent a few days in Moscow to see in situ the 
centre of the communist state). Nevertheless, according to a letter written in October 9 by 
Ehrenfest to Joff\'e,%
\footnote{Ehrenfest to Joff\'e, 9 October 1924, in \cite[pp.~171--172]{MoskovchenkoNI1990Ehrenfest}.}
we must suppose that those days Einstein was visiting Leiden (in fact, in the 
previously mentioned letter of 12 July, Einstein already announced a meeting at the ``beginning 
of October''). This is the excerpt of Ehrenfest's letter to Joff\'e that we are interested in:

\begin{quote}
My dear friend!

Precisely now Einstein is with us.
1.~We coincide fully with him that Bose's disgusting work by no means can be understood in the 
sense that Planck's radiation law agrees with light atoms moving independently (if they move 
\emph{independently} one of each other, the entropy of radiation would depend on the volume
\emph{not} as in Planck, but as in W.~Wien, i.e. in the following way:  $\kappa\log V^{E/h}$).\\

No, light atoms placed in the same cell of the phase space must depend one on the other 
in such a way that Planck's formula is obtained. Now we will clarify this question in a \emph{polemic} 
manner. I, Krutkow and Bursian will publish in the next number of Z. Physik a few considerations 
against, and simultaneously Einstein will give them answer in the same issue.%
\footnote{His emphasis.}
\end{quote}

Unless we are missing a letter, it is obvious from this document that Ehrenfest and Joff\'e had 
already discussed about Bose's ``disgusting'' work. This appellative can be understood as one of 
the first symptoms of Ehrenfest's future reluctance (which did not imply ignorance) towards the 
new mechanics; Ehrenfest did not hesitate to qualify it, some months later, as a ``sausage-%
machine-physics-mill.''%
\footnote{``Wurstmaschine-Physik-Betrieb,'' Ehrenfest to Einstein, 26 August 1926 (AEA 10-142). 
	English translation in \cite[p.~278]{MehraJEtAl1984Development4}.}

We have not been able to confirm whether Krutkow and Bursian---as the quoted letter 
suggests---went to Leiden with Ehrenfest on his return trip. We know that Krutkow enjoyed a 
scholarship from the Rockefeller foundation which allowed him to work in Western Europe in 
1925 and 1926, which he did for some time together with Born in G\"ottingen.\footnote{See \cite{FrenkelVI1990Physicists}.}
In any case, as far as we know, the ``considerations,'' which Ehrenfest referred to
in this letter were never published. 

It is very likely that Einstein referred to these debates in Leiden in the comment in his 1925 
paper in which he referred to Ehrenfest's objections. This does not mean that the ``others'' 
Einstein mentioned were the Russian friends of Ehrenfest. At least, the Austrian physicist Otto 
Halpern had pointed out to Einstein the lack of statistical independence of the molecules in the 
new approach.  He sent Einstein a detailed explanation of how the statistical independence of 
the elements under consideration had statistical implications. As he himself says, he based his 
reflections on Ehrenfest's and Krutkow's previous works. In his response, Einstein --who admits Halpern had ``illuminated very clearly a point of essential significance''--\footnote{\selectlanguage{german}
``Sie haben (...) einen Punkt von wesentlicher Bedeutung klar beleuchtet.'' Einstein to Halpern, September 1924 (AEA 12-128). Published
	in French translation in \cite[pp.~179--180]{OCAE01}.\label{note:Halpernletter}}distinguishes between two hypotheses:

\begin{quote}
1) All distributions of the individual quanta over the ``cells'' are equally probable (Wien's law).\\
2) All different quantum-distribution-pictures over the ``cells'' are equally probable (Planck's 
law). 
\end{quote}
And he continued:

\begin{quote}
Hypothesis 2 doesn't square with the hypothesis of the independent distribution of individual 
quanta---but expresses, in the language of the theory of existing quanta---a mutual dependence 
of the latter among each other.
 
Without experience one cannot decide between (1) and (2). The concept of independent
atom-like quanta calls for (1), but experience demands (2). Bose's derivation therefore cannot be 
regarded as a genuine theoretical justification of Planck's law, but only as a reduction of that law 
to a simple, but arbitrary statistical elementary hypothesis.
\end{quote}

Referring to his own extension of Bose's results to material gases, Einstein wrote:

\begin{quote}
This therefore also entails the implicit 
presupposition of certain statistical dependencies between the states of the molecules, a 
presupposition which the gas theory as such does not suggest. It would therefore be all the 
more interesting to know whether real gases behave according to this theory.
\footnote{\selectlanguage{german}
	``1) Alle Verteilungen der individuellen Quanten ueber die `Zellen' sind gleich wahrscheinlich
	(Wien'sches Gesetz).\\
	2) Alle verschiedenen Quanten-Verteilungs-Bilder ueber die Zellen sind gleich wahrscheinlich
	(Planck'sches Gesetz).\\
	Hypothese 2 passt nicht zur Hypothese der unabhaengigen Verteilung individueller Quanten,
	sondern drueckt---in der Sprache der Theorie existierender Quanten---eine gegenseitige
	Abhaengigkeit der letzteren von einander aus.\\
	Unabhaengig von der Erfahrung kann zwischen (1) und (2) nicht entschieden werden. Die
	Vorstellung unabhaengiger atomartiger Quanten verlangt (1), die Erfahrung jedoch velangt (2).
	Boses Ableitung kann also nicht als eine eigentliche theoretische Begruendung von Planck's
	Gesetz angesehen werden, sondern nur als dessen Zurueckfuehrung auf eine zwar einfache,
	aber willkuerlich statistische ElementarHypothese.\\
	(...) Es bedeutet dies also
	ebenfalls die implicite Voraussetzung gewisser statistischer Abhaengigkeiten zwischen
	den Zustaenden der Molekuele, fuer welche die Gastheorie als solche keine Anhaltspunkte
	liefert. Es waere also umso interessanter zu wissen, ob sich die wirklichen Gase gemaess
	dieser Theorie verhalten.'' Ibid.}
\end{quote}

Einstein's next visit to Leiden took place only in February of the following year, when he 
participated in the celebration of the fiftieth anniversary of Lorentz's doctorate.%
\footnote{Since that year the Royal Netherlands Academy of Arts and Sciences awards the Lorentz Medal.}  
On that occasion he spent only a few days in Leiden%
\footnote{See Einstein to Ehrenfest, 8 January 1925 (AEA 10-098): \selectlanguage{german}
	``Da ich aber im M\"arz nach
	Argentinien muss und hier in Berlin Vorlesung in diesem Semester halte, muss ich gleich
	wieder zur\"uck von Leiden.''}
and the prevailing agitation probably made it difficult for the  friends to discuss the matter calmly.%
\footnote{It was presumably during this short trip that Einstein gave Ehrenfest the manuscript of the
	second paper that was found in Ehrenfest's personal library, see HPE,
	Document EB22.} 

But a possible earlier meeting could have taken place in Berlin, where Ehrenfest spent some 
time in the beginning of November. This is suggested by a letter written by Ehrenfest, in which, 
however, he does not say anything related to the question of the gas.  But Einstein did comment 
on the subject in his next letter, dated on November 29, in which he mentioned the 
condensation phenomenon:

\begin{quote}
I am investigating the degeneracy function more thoroughly with Grommer. With a certain 
temperature the molecules ``condense'' without attractive forces, i.e., they pile up at the velocity 
zero. The theory is beautiful but does it also have some truth? I want to try whether one can 
also relate this to the dependence of the thermo forces at low temperature.%
\footnote{\selectlanguage{german}
	``Ich untersuche mit Grommer die Entartungsfunktion der Gase genauer. Von einer gewissen
	Temperatur an \glqq kondensieren\grqq die Molek\"ule ohne Anziehungskr\"afte, d.h.\ sie
	h\"aufen sich bei der Geschwindigkeit null. Die Theorie ist h\"ubsch, aber ob auch was
	Wahres dran ist? Ich will versuchen ob man den Verlauf der Thermokr\"afte bei tiefen
	Temperaturen damit in Zusammenhang bringen kann.'' Einstein to Ehrenfest,
	 29 November 1924 (AEA 10-093). \label{note:Grommer}}
\end{quote}

Everything seems to indicate that these days they hardly discussed. Einstein complained about 
that in another letter:

\begin{quote}
The thing with the quantum gas turns out to be very interesting. It seems to me more and more 
that something deep and true is hiding there. I am looking forward to---arguing about this with 
you.%
\footnote{\selectlanguage{german}
	``Die Sache mit dem Quantengas macht sich sehr interessant. Es kommt mir immer mehr vor,
	dass da viel Wahres und Tiefes dahinter steckt. Ich freue mich, bis wir dar\"uber -- streiten
	k\"onnen.'' Einstein to Ehrenfest, 2 December 1924 (AEA 10-095).}
\end{quote}

We guess they talked only after Einstein had presented the second paper of his theory in Berlin. 
Writing to confirm his participation at Lorentz's jubilee (and of his inability to stay longer than a 
few days), dated January 8 (precisely the day of his second presentation in the Academy), 
Einstein communicated to his friend that he had found a new way to justify the theory:

\begin{quote}
I will then completely convince you about the gas-degeneracy-equation, I found another sound if 
only not totally complete approach to it, free of the incriminated statistics. But how to set up a 
mechanics that leads to something like this? Presently I am plagueing myself roughly following 
Tetrode's, the invisible, prescription (Zeitschr.~f.~Physik 1922). There is something genial about 
this man.%
\footnote{\selectlanguage{german}
	``Ich werde Dich dann v\"ollig \"uberzeugen von der Gas-Entartungs-Gleichung, ich habe
	noch einen sicheren, aber allerdings nicht ganz vollst\"andigen Zugang zu ihr gefunden,
	frei von der inkriminierten Statistik. Aber wie die Mechanik aufstellen, die zu so was
	f\"uhrt? Gegenw\"artig plage ich mich ungef\"ahr nach dem Rezept Tetrodes
	(Zeitschr. f. Physik 1922), des Unsichtbaren. Es ist etwas Geniales an diesem Mann.''
	Einstein to Ehrenfest, 8 January 1925 (AEA 10-097). Einstein is referring to Tetrode's
	proposal of extending classical mechanics, which was published in 1922, see
	\cite{TetrodeH1922Wirkungszusammenhang}. \label{note:inkri}}
\end{quote}

Note the third instalment was read by Einstein on 29 January. According to this letter, we 
conclude that when he read the second one he had already thought about how to justify his 
theory with non-statistical arguments based on the displacement law. 
Note that up to this point neither Ehrenfest nor Einstein considered Bose's derivation a 
``theoretical foundation'' of Planck's law.

\section{The third, non-statistical paper}

On 29 January 1925, Einstein presented the third and last paper of his quantum theory of the 
monatomic ideal gas to the Prussian Academy for publication in its Proceedings.  This time, the 
sections and equations were not labelled consecutively with those of the preceding paper. 
Below we will return to this question in more detail, but these external aspects already suggest 
that the third contribution represents a path disconnected from the previous treatments. Or, at 
least, it seems Einstein wanted to present it this way.  In this section, we will paraphrase 
Einstein's arguments, closely following his original paper.

\subsection{Introduction and approach}

Einstein stated in the beginning that his theory was justified on the assumption that a light 
quantum differs, apart from polarization, from a material gas molecule only in the vanishing of 
its rest mass. This assumption was not taken for granted by many of his colleagues, nor had 
many researchers already accepted the statistical method used by Bose and by him. Einstein 
admitted that this method was ``not at all free of doubt'' (``keineswegs zweifelsfrei'')  and that it 
was only justified a posteriori by its success in the case of electromagnetic radiation.
Consequently, in the third paper, he was 
looking for new arguments in support of the new theory.

Nevertheless, this approach would 
follow his general heuristics of exploiting the gas-radiation analogy:

\begin{quote}
Here we plan to engage in considerations, in the field of gas theory, that are largely analogous, 
in method and outcome, to those that lead, in the field of radiation theory, to Wien's 
displacement law.%
\footnote{\selectlanguage{german}
	``Es handelt sich hier darum, auf dem Gebiete der Gastheorie Betrachtungen anzustellen,
	welche in Methode und Ergebnis weitgehend analog sind denjenigen, welche auf dem Gebiet
	der Strahlungstheorie zum {\scshape Wien}schen Verschiebungsgesetz f\"uhren.''
	\cite[p.~18]{EinsteinA1925Quantentheorie2}.}
\end{quote}

The results of these considerations will be restrictions on the form of the distribution function 
\begin{equation}
\rho = \rho(L,\kappa T,V,m)
\end{equation}
($L$ is the kinetic energy, $\kappa$ the Boltzmann constant, $T$ the temperature, $V$ the volume 
and $m$ the mass of the molecules). 

Einstein began by defining the subject of his investigation: consider a mole of an ideal gas 
contained in a volume $V$ at a temperature $T$ whose molecules have mass $m$. The distribution law 
will be of the form
\begin{equation}
dn = \rho(L,\kappa T,V,m)\frac{Vdp_1dp_2dp_3}{h^3}.
\label{eq:defdn}
\end{equation}
Here $dn$ designates---for fixed temperature, volume and mass---the number of molecules, whose 
Cartesian components of the momenta are in the range 
($p_1\dots p_1+dp_1$, $p_2\dots p_2+dp_2$, $p_3\dots p_3+dp_3$).  
Due to the isotropy of the problem, these components 
will appear in the argument of the distribution function $\rho$ only in the combination
\begin{equation}
L = \frac{1}{2m}\left(p_1^2+p_2^2+p_3^2\right),
\end{equation}
i.e., $\rho$ will depend only on the kinetic energy $L$. Within this approach, knowledge of
$\rho$ means knowledge of the equation of state, because  

\begin{quote}
... there can be no doubt that the pressure is obtained by mechanical calculation based on the 
collisions of the molecules with the wall.%
\footnote{\selectlanguage{german}
	``... nicht daran zu zweifeln ist, da{\ss} f\"ur den Druck die mechanische Berechnung
	aus den Zusammenst\"o{\ss}en der Molek\"ule mit der Wand ma{\ss}gebend ist.'' Ibid., 19. }
\end{quote}

By contrast, Einstein did not assume that collisions between molecules be governed by the laws 
of mechanics. He asserted that if that would be the case, one would arrive at Maxwell's 
distribution law and the classical equation of state. In fact, as we will see, he neglected 
interactions among molecules, as is appropriate for an ideal gas; we must add that it is not true 
that all kinds of mechanical interactions lead to Maxwell's distribution: we shall assume that 
Einstein was referring only to elastic collisions.

\subsection{Incompatibility with quantum theory}

Before beginning the analysis, Einstein commented again on Nernst's principle. However, he did 
not consider the delicate question of the factor $N!$ (required for $S$ to be an extensive quantity in classical theory), but this time he focused on the dependence 
of the entropy on volume which is contained in the additive term
\begin{equation}
\kappa\log V. 
\end{equation}

According to the Planckian formulation of Boltzmann's principle,
\begin{equation}
S=\kappa\log W,
\label{eq_S}
\end{equation}
and ever since Planck's first works on quantum theory, $W$ had always been taken to be a 
positive integer denoting the number of ways a certain macroscopic (thermodynamic) state can 
be built up microscopically. Therefore---Einstein added---, it made no sense if $W$ contained 
additive constants. In his opinion, if one takes into account Nernst's principle, the Planckian 
formulation (\ref{eq_S}) becomes almost a necessity: At absolute zero temperature thermal agitation 
ceases, and then there is only one possible microscopic configuration. That is, at absolute zero, 
one has:
\begin{equation}
W=1   \Longrightarrow   S=0 ,
\label{eq:W1S0}
\end{equation}
which precisely implies that Nernst's principle is satisfied.
Although he did not state it explicitly, it 
seems Einstein was refining his previous arguments on this subject. Remember that the 
extensivity of $S$ in the classical theory required the addition of a constant. In the new theory this 
is not the case.

What Einstein really wanted to emphasize is that this interpretation implies that the entropy 
cannot be negative. If we consider the Sackur-Tetrode equation of state of an ideal gas,%
\footnote{This expression does not appear in Einstein's paper, see \cite[p.~24]{PathriaRK2007Mechanics}.}
\begin{equation}
S=N\kappa\left\{\log\left(\frac{2\pi m\kappa T}{h^2}\right)^{\frac{3}{2}}\frac{V}{N} 
+ \frac{5}{2}\right\},
\end{equation}
we see that, if the volume is small enough, the entropy would be negative. Does this mean that 
real gases, contrary to what is implied by Nernst's theorem, can have negative entropies? No. It 
simply means that the classical theory of ideal gases can only be taken as valid under certain 
conditions. This is in analogy, so Einstein, to the case of   Wien's radiation law.

\subsection{Dimensional analysis}

According to its definition, (\ref{eq:defdn}), the distribution function $\rho$ is dimensionless.
Starting from that,
Einstein derived some of its properties assuming that Planck's constant, $h$, is the only 
dimensional constant contained in $\rho$ (he did not take into account Boltzmann's constant because 
he assumed that it would always appear in conjunction with temperature; in other words, he 
avoided selecting a specific temperature scale). Therefore, one can deduce, ``in a well known 
way,'' as Einstein remarked,%
\footnote{``in bekannter Weise,'' ibid., p.~20.}
that $\rho$ must be of the form:
\begin{equation}
\rho = \Psi\left( \frac{L}{\kappa T},
		\frac{m\left(\frac{V}{N}\right)^{\frac{2}{3}}\kappa T}{h^2}\right).
\label{eq:rhopsi2args}
\end{equation}

$\Psi$ is a universal, but unknown function, which depends on two dimensionless variables; it must 
satisfy the relation:
\begin{equation}
\frac{V}{h^3}\int\rho d\Phi=N,
\label{eq:intrhoN}
\end{equation}
where
\begin{equation}
d\Phi = \int_L^{L+dL} dp_1dp_2dp_3 = 2\pi (2m)^{\frac{3}{2}}L^{\frac{1}{2}}dL
\label{eq:dPhiint}
\end{equation}
(this is the same expression as (\ref{eq:densityofstates}) in Einstein's first paper, but
 without a contribution from 
position coordinates and without division by $h^3$). Nothing else can be said about $\rho$ 
on the basis of dimensional analysis alone. But the number of arguments for $\rho$ can be reduced 
from two to one variable, without introducing additional ``questionable'' assumptions. Einstein 
proposed two of those: 

\begin{enumerate}
\item The entropy of an ideal gas does not change in an ``infinitely slow adiabatic'' (sic) 
compression.%
\footnote{``bei unendlich langsamer adiabatischer Kompression.'' Ibid., p.~20.}
\item The required velocity distribution is valid for an ideal gas also in an external field of 
conservative forces. 
\end{enumerate}

Einstein argued that these two properties should be valid disregarding collisions. But the neglect 
of intermolecular collisions made their assumption unprovable, even if they would be ``very 
natural.'' In support of both, he announced they would lead not only to the same result, but also 
to a result according to which Maxwell's distribution law is valid in the region where quantum 
effects can be neglected. 

\subsection{Adiabatic compression} 

Einstein considered a gas with isotropic velocity distribution confined in a parallelepipedal 
container with sides of length $l_1$, $l_2$, and $l_3$. Since collisions of the molecules against 
the walls are elastic, they do not change the velocity distribution, which is stationary. 
The distribution has  the form:
\begin{equation}
dn = \frac{V}{h^3}\rho d\Phi,
\label{eq:dnform}
\end{equation}
where $\rho$ is an arbitrary function of the kinetic energy $L$. 
Although he did not state it explicitly, 
Einstein did not take up the result here that he had derived using dimensional 
analysis---(\ref{eq:rhopsi2args})---in 
the previous section. He simply started out from an isotropic distribution law. 
For an (infinitesimal) adiabatic compression that satisfies
\begin{equation}
\frac{\Delta l_1}{l_1} = 
\frac{\Delta l_2}{l_2} = 
\frac{\Delta l_3}{l_3} = 
\frac{1}{3}\frac{\Delta V}{V}.
\end{equation}
(the symbol $\Delta$ indicates that the process is adiabatic),
the distribution will stay of the form (\ref{eq:dnform}),  
but what will it be? The kinetic energy variation will be:%
\footnote{Einstein wrote $\delta$ instead of $\Delta$ for the variation of $p_i$
	in this equation. We think this is a typographical mistake. See \cite[p.~21]{EinsteinA1925Quantentheorie2}.}
\begin{equation}
\Delta L = \frac{1}{m}\left(|p_1|\Delta|p_1|+\dots+\dots\right) 
	= -\frac{2}{3}L\frac{\Delta V}{V}.
\label{eq:DeltaL}
\end{equation}

Since, according to (\ref{eq:W1S0}), we have
\begin{equation}
\Delta d\Phi = 2\pi (2m)^{\frac{3}{2}}\left(L^{\frac{1}{2}}\Delta dL
		+ \frac{1}{2}L^{-\frac{1}{2}}\Delta LdL\right),
\end{equation}
it follows that:
\begin{equation}
\Delta(Vd\Phi)=0.
\label{eq:DVdPhi}
\end{equation}

Since an adiabatic transformation does not change the number of molecules, it can be easily 
seen that:
\begin{equation}
\Delta\rho = 0.
\label{eq:drho0}
\end{equation}

As to the entropy, Einstein surmised that it would be of the form:
\begin{equation}
\frac{dS}{\kappa} = \frac{V}{h^3}s(\rho,L)d\Phi,
\label{eq:dSoverk}
\end{equation}
where $s$ is an unknown function. In adiabatic processes one has:
\begin{equation}
\Delta dS=0,
\end{equation}
and, therefore, according to (\ref{eq:DVdPhi}):
\begin{equation}
0 = \Delta s = \frac{\partial s}{\partial \rho}\Delta\rho + \frac{\partial s}{\partial L}\Delta L.
\end{equation}
Finally, using (\ref{eq:drho0}), it follows that $s$ is only a function of
$\rho$ alone,	
\begin{equation}
s=s(\rho).
\end{equation}

In a next step, Einstein imposed the condition that in thermodynamic equilibrium the 
entropy is a maximum with respect to variations of $\rho$, 
keeping fixed the number of particles and the total energy. The calculation gives:
\begin{equation}
\frac{\partial s}{\partial \rho} = AL+B,
\end{equation}
where $A$ and $B$ do not depend on $L$. Since $s$ only depends on $\rho$, its derivative will 
do so, too, and Einstein could write:
\begin{equation}
\rho = \Psi(AL+B).
\label{eq:PsiALB}
\end{equation}

In order to determine $A$, Einstein now considered the process of an infinitesimal isopycnic 
warming, i.e.\ a warming that does not alter the density of molecules (a transformation of 
constant volume in this case). Let $\mathcal{E}$ be the energy of the gas and let $D$ symbolize
this process, 
then
\begin{equation}
D\mathcal{E} = \frac{V}{h^3}\int L(D\rho)d\Phi
\end{equation}
will be equal---according to the thermodynamic relation between entropy and energy---to:
\begin{equation}
TDS = \frac{V\kappa T}{h^3} \int (Ds)d\Phi .
\end{equation}

Applying the chain rule and taking into account the invariance of the total number of molecules 
Einstein obtained:
\begin{equation}
\int L(D\rho)(1-\kappa TA)d\Phi = 0,
\end{equation}
that is:
\begin{equation}
A = \frac{1}{\kappa T}.
\end{equation}

Returning to expression (\ref{eq:PsiALB}), the functional dependence of the energy density with kinetic 
energy is:
\begin{equation}
\rho = \Psi\left(\frac{L}{\kappa T} + B\right).
\label{eq:rhopsi}
\end{equation}

\subsection{Ideal gas in a field of conservative forces}

The other path taken by Einstein with the aim of reducing the number of arguments for the 
distribution function appeals to its stationarity. Consider a gas in dynamic equilibrium in an 
external field of conservative forces whose single-particle potential energy $\Pi$ depends on the 
position of molecules. In addition to neglecting the collisions between them, Einstein assumed 
that ``the motion of the individual molecules under the influence of the external field follows 
classical mechanics.'' The condition, which $\rho$ must satisfy to be stationary, is:
\begin{equation}
\sum_i \left(\frac{\partial(\rho\dot{x}_i)}{\partial x_i} 
	+ \frac{\partial(\rho \dot{p}_i)}{\partial p_i}\right) = 0
\end{equation}
($x_i$ are the position variables). If we make use of the equations of motion the stationarity 
condition reads:
\begin{equation}
\frac{\partial \rho}{\partial x_i} \dot{x}_i + \frac{\partial \rho}{\partial p_i}\dot{p}_i=0.
\end{equation}

It follows that $\rho$ is constant along phase trajectories. Since, as a consequence of the problem's 
isotropy, the momenta $p_i$ will only appear in the argument in the combination of the kinetic 
energy $L$, $\rho$ must be of the form:
\begin{equation}
\rho = \Psi^{\star}(L + \Pi),
\label{eq:rhopsistar}
\end{equation}
with $\Psi^{\star}$, again, a universal, but unknown function.

Hence, the volume dependence on the distribution function will only come about through $\Pi$.

\subsection{Conclusions}

The results of the previous sections, (\ref{eq:rhopsi}) and (\ref{eq:rhopsistar}), are:
\begin{equation}
\rho = \Psi\left( h, m, \frac{L}{\kappa T} + B\right)
\end{equation}
\begin{equation}
\rho = \Psi^{\star}\left( h, m, \kappa T, L + \Pi\right)
\end{equation}
$B$ and $\Pi$ are universal functions that depend on $h$, $m$, $\kappa T$ and $V$. 
$\Psi$ and $\Psi^{\star}$ are universal and 
dimensionless functions. Taking into account also (\ref{eq:rhopsi2args}), Einstein derived:
\begin{equation}
\rho = \Psi \left( \frac{L}{\kappa T} +
	\chi\left(\frac{m\left(\frac{V}{N}\right)^{\frac{2}{3}}\kappa T}{h^2}\right)\right).
\label{eq:rhoresult}
\end{equation}

Here $\Psi$ and $\chi$ are universal functions of dimensionless variables. Both are related by equations 
(\ref{eq:intrhoN})---the number of particles---and (\ref{eq:dPhiint})---the density of states---,
and the problem was therefore reduced to finding the function $\Psi$. 

In the last paragraph, Einstein looked at the case in which the constant $h$ disappears from 
$dn$, i.e.\ at the classical limit. He defined:
\begin{equation}
u = \frac{Nh^3}{(m\kappa T)^{\frac{3}{2}}V} \qquad\text{and}\qquad v= \frac{L}{\kappa T}.
\label{eq:defuv}
\end{equation}

From (\ref{eq:defdn}) and (\ref{eq:defuv}) we see that $h$ will disappear if and only if
$\frac{1}{u}\psi$
does not depend on $u$. In this case, let
$\overline{\psi}(v)$
be the resulting function. With a suitable choice of the function $\chi$ 
(denoted in this particular case by $\phi$) it can be achieved that:
\begin{equation}
\psi(v+\phi(u)) = u\overline{\psi}(v).
\end{equation}

Differentiating first with respect to $u$ and then with respect to $v$, it can be seen that $\log\psi$ 
must be a linear function. It is then easy to see that $\psi$ must be the exponential distribution law,
i.e.\  Maxwell-Boltzmann's law:%
\footnote{Although in the paper this is not the case, we think that this expression and the following
	must have a bar on the top, as we write. Equation (\ref{eq:psibarv2}) involves another 
	mistake. We will discuss this question below.}
\begin{equation}
\overline{\psi}(v) = e^{-v}.
\label{eq:psibarv}
\end{equation}
In contrast, Einstein's statistical theory had produced the expression:
\begin{equation}
\overline{\psi}(v) = \frac{1}{e^v-1}.
\label{eq:psibarv2}
\end{equation}

Summarizing, Einstein pointed out that two aims have been achieved: 

\begin{quote}
First, we found a general condition (equation [\ref{eq:rhoresult}]), which has to be satisfied by any theory of 
the ideal gas. Second, it follows from the above that the equation of state which I derived will not 
be changed by either adiabatic compression or by the existence of conservative force fields.%
\footnote{\selectlanguage{german}
``Erstens ist eine allgemeine Bedingung (Gleichung (18)) gefunden worden, der jede
Theorie des idealen Gases gen\"ugen mu{\ss}. Zweitens geht aus dem Obigen hervor, da{\ss} die
von mir abgeleitete Zustandsgleichung durch adiabatische Kompression sowie durch konservative
Kraftfleder nicht gest\"ort wird.'' \cite[p.~25]{EinsteinA1925Quantentheorie2}.\label{note:lastpara}}
\end{quote}

\section{A displacement law for gases}

The novelty of Bose's approach was its statistical (microscopic) reasoning. At a thermodynamic 
(macroscopic) level, Bose did not obtain any new results, since it was precisely Planck's 
radiation law that was being derived. By contrast, for ideal gases there existed no such 
distribution law that had to be derived, only a distribution law that was valid in the classical limit. 
Schematically, the situation was as follows:

\bigskip
\begin{tabular}{|p{2cm}|p{4cm}|p{4cm}|}
\hline 
$ $	& Combinatorics & Thermodynamics-Statistical Distribution 
($\rightarrow$ \textit{indicates the classical limit})\\ 
\hline
Radiation & Bose & Planck's radiation law $\rightarrow$ Rayleigh-Jeans' (Wien's displacement law is always valid)\\
\hline
Gas & Bose-Einstein & Einstein's distribution law $\rightarrow$ Maxwell-Boltzmann's distribution law\\
\hline
\end{tabular}
\bigskip

	Einstein's theory thus implied developments at both levels. Indeed, the situation was 
similar to what had taken place some twenty-five years ago in the study of electromagnetic 
radiation. At 
the end of the nineteenth century, neither the radiation law nor the mechanism that is 
responsible for producing the thermodynamic equilibrium were known or, more precisely, they 
were known but had led to both empirically and theoretically wrong results. However, at the 
macroscopic level there was a guide post: Wien's displacement law. It restricts the arguments of 
the radiation law:
\begin{equation}
r(\nu, T)d\nu = \frac{8\pi \nu^2}{c^3}f\left(\frac{\nu}{T}\right)d\nu.
\label{eq:Wiendis}
\end{equation}

Therefore, once the spectrum is known at a certain temperature, it may be extrapolated to other 
temperatures. The derivations of Wien's law---there were several%
\footnote{See, for instance, \cite{WienW1894Temperatur} and \cite{LorentzHA1901Laws}.}
---always made use, at some 
point, of the second law of thermodynamics and, specifically, of the connection between states 
of equilibrium that were related by an adiabatic compression of the radiating cavity, i.e., by an 
infinitely slow compression in which the work transforms completely into internal energy. Max 
Planck---who also gave a demonstration in his \emph{Lectures on Heat Radiation}%
\footnote{\cite[pp.~314--332]{PlanckM1988Theory}.}
---justified the form of 
the quantum $h\nu$ appealing to that law.%
\footnote{See \cite{PlanckM1900Theorie}. In \cite[vol.~1, p.~703]{PlanckM1958Abhandlungen}.}

It is therefore not surprising that Einstein looked for analogous guide posts for derivations of the 
new distribution law. Nor it is surprising that he used an adiabatic transformation. He himself 
had suggested in his famous paper of 1905 on energy quanta that one could use an adiabatic 
compression to reduce the argument of the spectral entropy density $\varphi(r,\nu)$, which was defined 
as follows: 
\begin{equation}
S = V\int_0^{\infty} \varphi(r,\nu)d\nu
\label{eq:defS}
\end{equation}
($S$ is the entropy of radiation, $r$ the density of radiant energy):
	
\begin{quote}
One can reduce $\varphi$ to a function of a single variable by formulating the assertion that adiabatic 
compression of radiation between reflecting walls does not change its entropy. However, we 
shall not enter into this, but will immediately investigate how the function $\varphi$ can be 
obtained from  the black-body radiation law.%
\footnote{\selectlanguage{german}
	``Es kann $\varphi$ auf eine Funktion von nur einer Variabeln reduziert werden durch 
	Formulierung der Aussage, dass durch adiabatische Kompression einer Strahlung zwischen 
	spiegelnden W\"anden, deren Entropie nicht ge\"andert wird. Wir wollen jedoch hierauf nicht
	eintreten, sondern gleich untersuchen, wie die Funktion $\varphi$ aus dem 
	Strahlungsgesetz des schwarzen K\"orpers ermittelt werden kann.'' Ibid.}
\end{quote}
In 1905 Einstein did not elaborate on the argument because he was interested in other 
properties of $\varphi$. Probably, he did not dwell on this result because it was well known by his colleagues. One can show that:

\begin{equation}
\varphi(\rho,\nu)=\frac{8\pi V}{c^3} {\nu}^2 \xi\left(\frac{r}{\nu^3}\right),
\end{equation}
which is another way of enunciating Wien's displacement law. In 1925 Einstein may have recalled his own procedure of twenty years earlier. 

In our opinion, it was Paul Ehrenfest who gave the most detailed analysis of the radiation law 
and its derivations. Although his analysis was published over various articles, here we want to refer 
specifically to a paper of 1911, which we have discussed in more detail in other places.%
\footnote{\cite{EhrenfestP1911Zuege}. See \cite{NavarroLEtal2004Ehrenfest,NavarroLEtal2006Ehrenfest}.}
It will be 
useful to bring it up again here in order to comment on Einstein's non-statistical paper.	

	In his paper, Ehrenfest set out by listing the conditions that the radiation law necessarily 
needs to satisfy. Since these conditions comprise the quantum aspects of the radiation law, 
analogous conditions should hold for the new distribution law $\rho$ that Einstein was trying to justify. 
Ehrenfest's first three conditions were: 

\begin{enumerate}
\item The entropy does not change in an adiabatic compression. 
\item The radiation law satisfies the displacement law. 
\item The classical limit is obtained in the region where $\nu/T$  is small.
\end{enumerate}

Three more conditions were related to the violet region (large $\nu/T$) and required the avoidance 
of the so-called ultraviolet catastrophe (a divergence in the total energy), and, on the other 
hand, expressed analytically the behavior of Wien's and Planck's radiation laws.

What did Einstein have at his disposal in 1925 to ensure the validity of his material gas distribution law? Initially, it 
was only the third condition, i.e., the fact that his law produced the Maxwell-Boltzmann's 
distribution law in the classical limit. But at \emph{low} temperatures (or \emph{high} densities) 
he did not have 
anything comparable to Wien's or Planck's radiation laws. In this region, only Nernst's theorem 
supported the expression for the entropy obtained by Einstein. 

Hence, conditions 1 and 2 were not available. As we have said before, in a certain sense, they 
are both related, and it should be noted here that Ehrenfest listed them separately because his 
analysis aimed at drawing conclusions at the microscopic level, where the (mechanical) 
adiabatic invariants play a crucial role. Einstein in 1925, in contrast, was not interested in 
mechanical invariants because his research remained at the thermodynamic level.
According to this scheme, the problem of electromagnetic radiation served Einstein as a guide 
to explore the case of the degenerate ideal gas.%
\footnote{The degenerate gas had been defined in opposition to a perfect gas by Walther Nernst. As far as we know, it was Nernst who first introduced this terminology, referring to gases at low temperatures, in which their translational contribution to specific heat tended to disappear, see \cite[p.~493]{DesalvoA1992Constant}}
He justified the additivity of the entropy (\ref{eq:dSoverk}) with respect to those 
portions of gas with different kinetic energy $dL$, as follows:

\begin{quote}
This hypothesis is analogous to the one used in radiation theory, according to which the entropy 
of the radiation is composed additively from quasi-monochromatic parts. It is equivalent to the 
assumption that one may introduce semi-permeable walls for molecules of different ranges of 
velocity.%
\footnote{\selectlanguage{german}
	``Diese Hypothese ist in der Strahlungstheorie jener analog, da{\ss} die Entropie
	einer Strahlung sich aus der der quasi-monochromatischen Bestandteile additiv
	zusammensetze. Sie ist \"aquivalent der Annahme, da{\ss} man f\"ur Molek\"ule
	verschiedener Geschwindigkeitsbereiche semi-permeable W\"ande einf\"uhren d\"urfe.''
	\cite[p.~21]{EinsteinA1925Quantentheorie2}.}
\end{quote}

This is consistent with his justification of writing the total entropy as an integral over the spectral 
entropy density (equation (\ref{eq:defS}) above) in 1905 (the analogue to eq.~(\ref{eq:dSoverk})):

\begin{quote}
... radiations of different frequencies are to be viewed as separable from each other without 
expenditure of work and without supply of heat ... %
\footnote{\selectlanguage{german}
	``[...] Strahlungen von verschiedenen Frequenzen [sind] als ohne Arbeitsleistung und ohne
	W\"armezufuhr voneinander trennbar anzusehen [...]'' \cite[p.~137]{Einstein1905i}.}
\end{quote}

It is curious to see how in Ehrenfest's 1911 paper the direction of the justification was just the 
opposite than in Einstein's 1925 paper. His approach was inspired by gas theory:

\begin{quote}
We determine, for given total energy, the ``most probable'' distribution of oscillations over all 
possible ranges of excitations according to the same procedure that Boltzmann had used to 
determine, for given total energy the ``most probable'' distribution of molecules for a mixture of 
gases consisting of many kinds of molecules over all possible ranges of velocity. The eigen 
frequencies of one and the same frequency range $d\nu$ here play the same role as the molecules 
of one and the same substance.%
\footnote{\selectlanguage{german}
	``Es wird bei gegebener Totalenergie die \glqq wahrscheinlichste\grqq ~Verteilung
	der Eigenschwingungen \"uber alle m\"oglichen Erregungsbereiche nach demselben
	Verfahren bestimmt, nach welchem {\scshape Boltzmann} die---bei gegebener
	Totalenergie---\glqq wahrscheinlichste\grqq~Verteilung der Molek\"ule eines aus vielen
	Molek\"ulsorten bestehenden Gasgemisches \"uber alle m\"oglichen 
	Geschwindigkeitsbereiche bestimmte. Die Eigenschwingungen eines und desselben 
	Frequenzbereiches $d\nu$ spielen dabei die Rolle der Molek\"ule einer und 
	derselben Substanz.'' \cite[pp.~94--95]{EhrenfestP1911Zuege}.}
\end{quote}

In ten years the reference system and the unknown system switched their roles. Such was the 
confusion into which quantum discoveries had brought physics at the beginning of the century.

Note that albeit both analogies are not the same they are equivalent. Einstein justifies expressions (\ref{eq:defS}) and (\ref{eq:dSoverk}) as follows. The fact that one can write the total entropy as a summation of all monochromatic or \textit{mono-kineticoenergetic} entropies means that total entropy is additive with respect to frequency (resp. kinetic energy). Ehrenfest, on the other hand, compares different frequencies with different substances, as the entropy of a mixture is also additive with respect to its components. In both cases entropy of radiation is the summation of monochromatic entropies.

The analogies invoked by Einstein do not end here. Although, as we have said, he used the 
idea of an adiabatic compression to derive something like a displacement law for ideal gases, 
the whole approach nevertheless rested on dimensional analysis. This approach was modelled 
on that in a 1909 paper on radiation, in which Einstein himself had given a derivation of the 
displacement law.

\subsection{Dimensional analysis}

Einstein had argued that the distribution function can depend only on two dimensionless 
quantities (cf.\ eq.~(\ref{eq:rhopsi2args})): 
\begin{equation}
\frac{L}{\kappa T} \qquad\text{and} \qquad 
\frac{m\left(\frac{V}{N}\right)^{\frac{2}{3}}\kappa T}{h^2}
\label{eq:2args}
\end{equation}
He did not present the arguments for his claim in any detail, but it is not very difficult to guess 
what they were. We must remember that Buckingham's theorem (the consequences of which, 
as we will argue, were certainly known to Einstein, if not by this name) states that the difference 
between the number of quantities that are assumed to play a role in the physical system under 
consideration, on the one hand, and the number of fundamental variables involved, on the other 
hand, gives the number of independent dimensionless monomials which can be constructed.%
\footnote{\cite{BuckinghamE1914Systems}. For historical accounts of dimensional analysis,
	see, e.g., \cite{BridgmanPW1922Analysis} and \cite{MagagnoE1971Review}.}
In our case the fundamental variables are mass ($M$), time ($T$) and length ($L$), and the quantities 
supposed by Einstein to be arguments of the distribution function $\rho$ are

\begin{itemize}
\item[$L$:] kinetic energy $[L]=ML^2T^{-2}$
\item[$\kappa T$:] temperature (multiplied by Boltzmann's constant), $[\kappa T]=ML^2T^{-2}$ 
\item[$V$:] volume, $[V]=L^3$
\item[$m$:] mass of the molecules, $[m]=M$
\end{itemize}

The choice of quantities that play a role in the problem is the crucial point in any dimensional 
analysis, since the result critically depends on it. Einstein chose the quantities that appear in the 
distribution law for the classical ideal gas. That is
\begin{align}
& dn_{\rm clas} = \frac{1}{V} \left(\frac{h^2}{2\pi m\kappa T}\right)^{\frac{3}{2}}
	e^{-\frac{L}{\kappa T}} \frac{Vdp_1dp_2dp_3}{h^3}\\
\Rightarrow &\rho_{\rm clas} = \frac{1}{V} \left(\frac{h^2}{2\pi m\kappa T}\right)^{\frac{3}{2}}
	e^{-\frac{L}{\kappa T}} = \rho(L,\kappa T, V, m). 
\label{eq:rhoclas}
\end{align}
Taking into account also Planck's constant $[h] = ML^2T^{-1}$---the distinguishing mark of any 
quantum phenomenon--%
\footnote{Of course, the constant $h$ did not appear in the classical expression for $dn$. However, once the phase space had been quantized, one could relate $h$ with the volume of a microstate. Only in this way can $h$  appear in the classical distribution function, (\ref{eq:rhoclas}). As in the statistical works prior to early quantum developments, what was significant was the quotient of the phase space volumes, the later appearance of $h$ did not retrospectively contradict  classical developments.} 
in the distribution function of the classical gas, albeit not in $dn$,
it follows from Buckingham's $\Pi$ theorem that only two independent dimensionless monomials can 
be constructed. One possible pair is the one that Einstein proposed, see eq.~(\ref{eq:2args}). But this 
choice is not unique. One could also have:
\begin{equation}
\frac{L}{\kappa T} \qquad \text{and} \qquad \frac{mV^{\frac{2}{3}}L}{h^2}
\end{equation}
or:
\begin{equation}
\frac{L}{\kappa T} \qquad \text{and} \qquad \frac{m^2V^{\frac{4}{3}}\kappa TL}{h^4}
\end{equation}
(Note that the number of particles $N$ does not play a role in these considerations, since it is 
dimensionless). In order to have only one monomial, one needs to impose an additional 
condition. Most likely, Einstein simply decided to choose the monomial that was compatible with 
the classical result (\ref{eq:rhoclas}).
It has the advantage of being a natural one: since the influence of the 
kinetic energy $L$ in the density function has to be weighted by the temperature $\kappa T$, both have to 
appear together in one of the two monomials. 

Once the first monomial has been chosen and after excluding from the second one the kinetic 
energy (for the reason we have explained), the number of possibilities reduces drastically. We 
now have a problem with four quantities (three quantities and one constant) and three 
fundamental variables. Thus, in this way, Einstein's result (\ref{eq:2args}) can be considered
univocal. Note that the two monomials (\ref{eq:2args}) can also be derived from the form
of the classical density (\ref{eq:rhoclas}).
Let's now jump to the end of Einstein's paper. After having reduced the number of arguments of 
the density function to a single parameter,
\begin{equation}
\frac{L}{\kappa T} + \chi \left( \frac{mV^{\frac{2}{3}}\kappa T}{h^2} \right),
\label{eq:singleparameter}
\end{equation}
Einstein showed that this result is in agreement also with other constraints. He looked 
at what happens if one assumes that Planck's constant does not appear in the expression for 
$dn$, and he obtained Maxwell-Boltzmann's distribution law. There is a mistake in testing that his 
theory also satisfies this dependence, but the mistake is inconsequential. Einstein wrote that the 
density can be written as
\begin{equation}
\frac{1}{e^{\frac{L}{\kappa T}}-1},
\end{equation}
which is not true. Nevertheless, his theory satisfies the conclusion of the dimensional argument, 
since $\rho$ is:
\begin{equation}
\frac{1}{\exp\left[{\frac{L}{\kappa T} + \chi \left( \frac{m(\frac{V}{N})^{\frac{2}{3}}\kappa T}{h^2} \right)}\right]-1}.
\end{equation}

Hence, his conclusion is correct but not the reasoning. 

To conclude, note that, in fact, the result achieved by Einstein in eq.~(\ref{eq:rhoresult})
does not lead to 
a true displacement law. This law was named this way because it implied that the maximum of 
intensity satisfied the relation
\begin{equation} 
\frac{\nu_{\rm max}}{T} = \text{const}.
\end{equation}
Therefore, the maximum displaces in proportionality with temperature. Due to the form of the 
argument in eq.~(\ref{eq:singleparameter}), this simple statement cannot be made in the case
of molecules. The 
simple state of affairs in the case of radiation is a consequence of the non-conservation of---or 
else the lack of sense of the concept of---the number of particles.

Let us now take a look at earlier considerations which may have inspired Einstein and which had 
helped him before when he was trying to find his way on slippery ground.

\subsection{Einstein's deduction of Wien's displacement's law in 1909.}

In January 1909, the editors of the Physikalische Zeitschrift received the manuscript of one of 
the subsequently more celebrated papers by Einstein. In it, and in the talk he gave in Salzburg 
in September of the same year, Einstein suggested and emphasized for the first time the essential 
dual nature of radiation---corpuscular and undulatory---, starting from a fluctuation analysis of 
radiation momentum.%
\footnote{\cite{Einstein1909b}. The presentation in Salzburg was transcribed in \cite{Einstein1909c}.}
This paper must be situated in a context where the physics community 
had not yet realized the implicit contradictions with classical physics inherent in Planck's 
radiation law.%
\footnote{\cite{KuhnT1978Theory}.}
It was only after 1911---after the first Solvay conference---when the position 
hitherto taken by a small minority would become the received opinion: That it would be 
necessary to undertake a deep revision of the existing physical theories in order to account for 
the quantization that Planck had introduced by his black body radiation theory.

In 1909, Einstein was one of the first physicists who had become clearly aware of the 
exceptional nature of the situation and suggested that perhaps the break did not have to be as 
traumatic as it may have seemed at first sight. He argued that one should relate to each other 
Planck's quanta and electricity quanta. The latter did not arise either in any natural way from 
Maxwellian electromagnetic theory, and maybe a modification of this theory could account for 
both instances of quantization. In support of the feasibility of this idea, Einstein invoked some 
dimensional considerations published by James H.~Jeans ``a few years ago.'' Although he did not 
cite the precise source, it is beyond doubt that Einstein referred to the paper ``On the laws of 
radiation,'' published by the British physicist in 1905.%  
\footnote{\cite{JeansJH1905Laws}. See \cite[p.~549, note~60]{CPAE02}, where the editors also 
	point out that, in contrast to Einstein's argument, Jeans did not use the Planck constant $h$,
	nor does he consider the ratio $e^2/c$.}

In it, Jeans gave a derivation of the displacement law, in which he outlined a dimensional 
argument that allowed him to write the constants that appear in this law and in Stefan's law 
depending on known universal constants. In other words, he excluded from his consideration 
Planck's constant. With this derivation, Jeans was trying to advance a new argument for his 
claim that the problem of specific heats and the black-body problem were both caused by the 
fact that thermodynamic equilibrium was not established. For this reason, the equipartition of 
energy could not be applied in the theoretical analysis of either system. In addition, he also 
argued for the electronic origin of radiation. 

Thus, Jeans proposed that electron trajectories are the source of the spectrum (and of its 
universality), taking up an old idea by Hendrik A.~Lorentz.%
\footnote{\cite{LorentzHA1903Emission}.}
According to Jeans, the radiant 
energy per volume at temperature $T$ and wavelength $\lambda$ depends on the following constants:%
\footnote{We omit the dependency on some ``specific constants'' of the body that do not play any role
	in what follows.}

\begin{itemize}
\item[$V$:] speed of light,
\item[$e$:] electron charge,
\item[$m$:] electron mass,
\item[$R$:] gas constant (the mean kinetic energy of 
 			  a single particle is $\frac{3}{2} RT$),
\item[$K:$] inductive capacity of the ether (Coulomb's law,              
                                $F=K^{-1}q_1q_2r^{-2}$).
\end{itemize}
As a dimensional basis, he took length $L$, mass $M$, time $t$, inductive capacity $K$, 
and temperature $T$:%
\footnote{As pointed out already in \cite{EhrenfestP1906Bemerkung}, there is a typographical error in
	\cite[p.~548]{JeansJH1905Laws}, who has the dimension of $R$ as $LMt^{-2}T^{-1}$.}

\begin{itemize}
\item[$\lambda$:]  $L$,
\item[$T$:]  $T$,
\item[$V$:]  $L t^{-1}$,
\item[$e$:]  $L^{3/2} M^{1/2} t^{-1} K^{1/2}$,
\item[$m$:]  $M$,
\item[$R$:]  $L^2 M t^{-2} T^{-1}$,
\item[$K$:]  $K$.
\end{itemize}
Since there are five dimensional units and seven quantities, Jeans was able to construct two 
independent dimensionless monomials. But which ones he chose, is one of the dark spots of his 
argument. The key step of his derivation was that one of the two monomials that he 
constructed, $c_1 = RTm^{-1}V^{-2}$, is of order $10^{-8}$ (at $100^{o}$C); 
he used this fact to justify  his claim 
that the radiation law depends only on the other monomial, which he chose as:
\begin{equation}
c_2 =  \frac{\lambda RTK}{e^2}.
\end{equation}

A few months later, Ehrenfest published a short note in the \emph{Physikalische Zeitschrift}, in 
which he criticized Jeans's argument.%
\footnote{\cite{EhrenfestP1906Bemerkung}.}
Indicating that he could not follow the reasoning at 
various steps, he focussed only on the arbitrariness of the choice of the pair of monomials and 
showed by means of a slightly different choice of monomials, ${c'}_1=c_1$, ${c'}_2= c_2c_1^{1/8}$, 
how Jeans's 
reasoning may lead also to a different result, i.e., not to the displacement law. The British 
physicist did not accept the criticism.%
\footnote{\cite{JeansJH1906Bemerrkung}.}
In his view, Ehrenfest's counterexample did not square 
with his proposal. Ehrenfest wrote another reply, defending himself against Jeans'
counter-attack and repeated his criticism of the dimensional argument.%
\footnote{\cite{EhrenfestP1906Bemerkung2}.}

In 1909, in his article ``Radiation Theory'' for the \emph{Encyklop\"adie der Mathematischen 
Wissenschaften}, Wilhelm Wien referred to Ehrenfest's refutation of Jeans's argument like this:

\begin{quote}
Another derivation of the displacement law is given by J.H.~Jeans. But there is an uncontrollable 
approximation in it, which must be introduced as a hypothetical assumption; therefore, Jeans's 
derivation cannot be regarded as a proof of the displacement law.%
\footnote{\selectlanguage{german}
	``Eine weitere Ableitung des Verschiebungsgesetzes gibt \emph{J.H.~Jeans}. Doch kommt in 
	ihr eine nicht kontrollierbare Vernachl\"assigung vor, welche als hypothetisch Annahme 
	einzuf\"uhren ist; daher kann die \emph{Jeans}sche Entwicklung nicht als Beweis f\"ur
	das Verschiebungsgesetz angesehen werden.'' \cite{WienW1909Theorie}.}
\end{quote}

Although we cannot give further evidence, it appears that Jeans finally admitted the weakness 
of his argument. His result may be understood as a piece of circumstantial evidence or as an 
illustration, not as a genuine derivation. Some years later, in his famous \emph{Report on Radiation 
and the Quantum-Theory} of 1914, Jeans had changed his attitude towards quantum theory and 
did not mention this result.%
\footnote{\cite{JeansJH1914Report}.}
What makes this omission even more significant is that Jeans here 
defended practically the same argument that Einstein gave in 1909---which we will discuss at 
once---, but without quoting him:

\begin{quote}
Any attempt to refer back the atomicity of $e$ to the structure of the ether simply discloses the fact 
that the fundamental equations of the ether are not yet fully known; it implies that if they were 
fully known they might be expected to contain the quantity $e$, and this is perhaps the same thing 
as saying that they would contain the quantity $h$. It may be that if the equations of the ether were 
fully known they would be seen to involve the quantum-theory.%
\footnote{Ibid., p.~81.}
\end{quote}

For Jeans the possibility of associating the atomicity of the charge with the quantum of energy 
was still alive in 1924, when he published the second edition of his \emph{Report}.%
\footnote{\cite{JeansJH1924Report}.}
The quoted text 
remained unchanged after a decade of further developments of quantum theory.

As we have said, Einstein took up Jeans's demonstration in his 1909 paper on the radiation 
problem but modified it at some points. The most important difference is that he did not assume 
that the radiation density would depend on the electron mass. This difference renders 
Ehrenfest's objections invalid, because now there is, according to the $\Pi$ theorem, only one 
possible dimensionless monomial. (This does not mean Einstein was aware of the controversy 
that his colleagues had maintained; we have not found any evidence in this respect.%
\footnote{See also Einstein to H.A.~Lorentz, 30 March 1909 \cite[Doc.~146]{CPAE05}, where 
	Einstein also refers to Jeans's argument. Einstein and Ehrenfest met in person only 
	a few years later, in 1912, see \cite[chap.~12]{KleinM1985Ehrenfest}, and also 
	\cite[pp.~172--175]{SauerT2007Einstein}.}) 

 In a cavity filled with gas molecules, radiation, and ions---the latter allow the energy exchange 
between the former---, the quantities that---according to Einstein---should be included as 
arguments of the spectral density are:

\begin{itemize}
\item[$RT/N$:] energy of a molecule (dimensionally speaking),
\item[$c$:] speed of light,
\item[$e$:] quantum of electricity,
\item[$\nu$:] frequency.
\end{itemize}

Only attending to the dimensions of the density of radiant energy $r$, which are%
\footnote{Einstein wrote $\rho$ instead of $r$ and $\epsilon$ instead of $e$. Now 
	$[e] = L^{3/2}M^{1/2}T^{-1}$.} 
$ML^{-1}T^{-1}$,
it can be seen that $r$ must have the form:
\begin{equation}
r = \frac{e^2}{c^4} \nu^3 \Psi \left(\frac{Ne^2}{Rc}\frac{\nu}{T}\right).
\label{eq:formofr}
\end{equation}
This is the only possible combination to establish a dimensionless relation for $r$ with the 
quantities considered by Einstein. Expression (\ref{eq:formofr}) is none other than Wien's
displacement law, cp.\ eq.~(\ref{eq:Wiendis}). Comparing now this result with Planck's radiation law,
\begin{equation}
r = \frac{\alpha\nu^2}{c^3}h\nu\frac{1}{e^{\frac{h\nu}{\kappa T}}-1}
\end{equation}
($\alpha$ is a dimensionless factor, cp.\ eq.~{\ref{eq:prefactor}), Einstein arrives at:
\begin{equation}
h=\frac{e^2}{c} \qquad \text{and} \qquad \kappa = \frac{R}{N}.
\end{equation}

He then observed that the first relation differs from the known results by three orders of 
magnitude, which in his opinion can be attributed to dimensionless factors. Similar to the way 
Jeans did it, he speculated about the possibility of reducing the ``light quantum constant $h$'' to the 
``elementary quantum of electricity $e$'' and thus about the unnecessariness of introducing new 
universal constants.

In a letter Einstein wrote to Lorentz in 6 May 1909, he regarded it as highly significant that the 
displacement law could be obtained through dimensional considerations, and insisted again on 
the existence of a relation between $e$ and $h$.%
\footnote{A.~Einstein to H.A.~Lorentz, 30 March 1909 \cite[Doc.~163]{CPAE05}.}
But after reading Einstein's paper, Lorentz 
responded that he did not regard a discrepancy of three orders of magnitude an insignificant 
one nor did he agree that the radiation law, which should give evidence of the properties of the 
ether, should include the electronic charge.%
\footnote{H.A.~Lorentz to A.~Einstein, 6 May 1909 \cite[Doc.~153]{CPAE05}.}
But that was precisely Einstein's bet in that state of 
ignorance and confusion about quantum phenomena.	

Thus, in his 1909 analysis of the radiation problem dimensional analysis provided an argument 
for strengthening a thesis shored up by Einstein in conjunction with other fundamental 
arguments, such as fluctuation analysis. But it appears that Lorentz, whom he admired deeply, 
had convinced Einstein of the weakness of the dimensional argument to the effect that he forgot 
about it. A little more than two years later, just before attending the first Solvay conference, in 
autumn 1911, Einstein responded to a question by Michele Besso as to whether he had ever 
come across a situation where one had to choose between the quantum of action and $e^2$ in 
order to introduce ``natural units.'' Einstein responded that one knew that the ratio was a factor 
of $900$ but that he had never come across this in dimensional considerations, as far as he 
remembered.%
\footnote{Einstein to Michele Besso, 11 September 1911, in \cite[Doc.~283]{CPAE05}. The
	reference to what it is that Einstein did not remember is unclear in the original 
	German. It could be the factor $900$, but more likely it is a response to Besso's
	question: ``Der Unterschied zwischen $e^2$ und $h$ ist ja Faktor $900$. Ist mir noch nicht
	bei Dimensionalbetrachtungen begegnet, soviel ich mich erinnere.''}
And in Brussels, he spent no time on this question in his presentation, although 
the relation between $e$ and $h$ appeared more than once during the meeting.%
\footnote{\cite[pp.~75--76, 131]{LangevinP1912Theorie}.}
In 1916 Arnold 
Sommerfeld introduced in the quantum debates the constant of fine structure,
\begin{equation}
\alpha = \frac{2\pi e^2}{hc},
\end{equation}
which contains, in a sense, the numerical relation among $e$, $h$ and $c$.%
\footnote{In the late forties, Pauli concluded his contribution to the Schilpp volume
	\emph{Albert Einstein Philosopher-Scientist} with a comment on Einstein's 1909 
	paper. He observed that ``the present form of quantum mechanics is far from
	anything final, but, on the contrary, leaves problems open which Einstein considered
	long ago.'' In the 1909 paper, Pauli continued, Einstein ``stresses the importance
	of Jeans' remark that the elementary charge $e$, with the help of the velocity of light
	$c$, determines the constant $e^2/c$ of the same dimension as the quantum of action $h$
	(thus aiming at the now well known fine structure constant $2\pi e^2/hc$).'' But the development
	of physics had not produced an understanding of the elementary charge flowing from a
	quantum theory. The determination of the fine structure constant therefore, was ``certainly
	the most important unsolved problem of modern physics.'' \cite[p.~158]{SchilppPA1949Einstein}.}

Just before going to Brussels to attend the Solvay meeting, Einstein appealed again to 
dimensional analysis in a paper that was not directly related to radiation. In it, he admitted that, 
in general, discrepancies in numerical values should be of the order of unity. However, he 
admitted exceptions.

\subsection{Einstein on the quantum theory of solids}

In 1907, Einstein had been a pioneer in applying quantum theory to the study of solids. 
Following his seminal work, Walther Nernst and his collaborators in Berlin demonstrated the 
wisdom of Einstein's speculations, at least in a broad sense. However, in 1911, it was clear that 
the theoretical curve deviated from the experimentally observed data at very low temperatures. 
Einstein then wrote two new papers with an attempt to leave aside the monochromatic normal 
modes that had characterized his first approach to the problem.%
\footnote{\cite{Einstein1911b}, \cite{Einstein1911g}. These papers are discussed in
        \cite{BridgmanPW1922Analysis}.}
(Shortly before, Nernst and 
Lindemann had proposed an alternative formula, which Einstein quoted in his 1911 papers.) 
These new developments in the field of crystalline solids culminated in 1912 with the 
appearance of the famous contributions by Peter Debye, on the one hand, and by Max Born 
and Theodor von K\'arm\'an, on the other. 

In the second paper of 1911, Einstein tried to argue that solids must present a set of 
frequencies related to the coupling among different forced motions of the atoms. In the third 
and fourth paragraphs we find two other instances of dimensional analysis. 

Here Einstein used it to derive an expression for the proper frequency of an atom in a solid, 
which he had given already in his earlier paper. He obtained a satisfactory result, since the 
dimensionless coefficient, which still had to be determined, is of the order of unity (both in the 
case of his formula and in the case of the Nernst-Lindemann formula). In addition, he used the 
opportunity to show that Lindemann's formula for the melting temperature of a solid also is in 
agreement with dimensional arguments.

In the last section, Einstein tried to find an expression for the thermal conductivity $K$. Using the 
dimensional method described earlier in the same paper, he arrives at the following functional 
dependence:%
\footnote{Einstein wrote $\tau$ instead of $\kappa T$.}
\begin{equation}
K = C\frac{\nu}{d}\varphi\left(\frac{md^2\nu^2}{\kappa T}\right)
\end{equation}
($m$ is the mass of an atom, $d$ the interatomic distance, $\nu$ the oscillation frequency and $C$ a 
constant). In order to determine the function $\varphi$ Einstein appealed to recently published 
measurements by Arnold Eucken, which indicated a dependency of $K$ with the inverse of 
temperature. Accordingly, the final expression should be:
\begin{equation}
K = C\frac{md\nu^3}{\kappa T}.
\end{equation}
This combination of dimensional analysis with an empirical law (referred only to one of the 
involved quantities) represents, in our opinion, an interesting example of how fruitful this 
procedure can be. 

In this section we have shown another example---the third---in which Einstein used a suitable 
analyzing method for exploring new territories. Therefore, he used dimensional analysis in 
radiation, solids, and gases.

We conclude our analysis of Einstein's use of the method of dimensional analysis with a 
comment on Tatiana Ehrenfest, who, in fact, published a number of papers on this problem. 
Indeed, in August 1925 she signed a paper, which contains explicit criticism of Einstein's use of 
dimensional analysis in 1909 and which was communicated to the \emph{Philosophical Magazine} by 
her husband. In  her paper, Tatiana Ehrenfest juxtaposed the method of dimensional analysis 
with what she called the ``theory of similitudes,'' the latter being based on a mathematical 
analysis of transformation properties of differential equations. In her comparison, she criticised 
the use of dimensional analysis in physics as being often misunderstood in its deductive power. 
The theory of similitudes, in contrast, she commended for being capable of producing definite 
and reliable results, provided two rules were followed. According to these rules, one needs to 
consider all ``fundamental equations'' and one must not introduce ``conditional equations'' except 
those that follow from the transformation properties of the fundamental equations. In her 
concluding paragraph, she attributed value to the method of dimensional analysis only in the 
case that the theory of similitude is not applicable:

\begin{quote}
However, if the fundamental equations of the two problem are unknown, of the two methods 
there remains only dimensional analysis. It must never be forgotten that in such cases one 
advances only gropingly, and without experimental or theoretical proof from another quarter one 
can never be completely certain of the results. Dimensional analysis combined with proof of this 
kind may be viewed as a systematic method for determining whether in the given problem new 
and unknown fundamental equations take part which are non-homogeneous relative to the 
quantitites considered; i.e. which involve dimensional coefficients or new variables.%
\footnote{\cite{EhrenfestT1926Analysis}.}
\end{quote}

Interestingly, she added a footnote here at the end of her paper, in which she referred to 
Einstein's 1909 paper as providing ``a pretty example of such an application of dimensional 
analysis.'' However, the reference is not to the example that we have discussed above, but to 
another instance in this paper, in which Einstein used a ``simple dimensional consideration.''%
\footnote{``einfache Dimensionalbetrachtung,'' \cite[p.~189]{Einstein1909b}.}
In this example, he argued that according to a dimensional argument the mean squared energy 
fluctuation $\overline{\epsilon^2}$ comes out non-classically.
It appears likely that she added the reference to 
Einstein's paper as an afterthought. Despite the praise implicit in the final footnote, she 
referenced the paper  once more in a footnote that she added to a statement where she pointed 
out that by ``erroneously [...] overlooking one or another of the fundamental equations" or else 
by ``forgetting'' interdependencies of the variables, ``there may be projected an illusory 
definiteness of solution.'' The footnote here says: ``This is the case in the example analysed by 
Einstein. Footnote of the final paragraph.''%
\footnote{Ibid., p.~266.}

We may extend Tatiana Ehrenfest's praise and criticism of Einstein's use of dimensional 
analysis in 1909 to the example we have discussed above. In an explicit example that she 
discussed in her paper, she criticized Lord Rayleigh to have used an unwarranted additional 
conditional equation, i.e., relating temperature to the average molecular kinetic energy, in 
violation of the second rule for proper use of the theory of similitudes: ``The equation by which 
temperature is defined as the average kinetic energy of the molecules is not one of the 
fundamental equations of the problem; it relates quantities (molecular velocity and molecular 
mass) which do not occur in any of the fundamental equations.''%
\footnote{Ibid., pp.~268--269.}

It is quite possible that Tatiana Ehrenfest added the reference to Einstein's 1909 paper at the 
suggestion of her husband. In any case, we note that the Ehrenfests had a sophisticated 
understanding of the intricacies of the dimensional analysis and had well-informed and critical 
opinions on the status of the results based on dimensional considerations, including those put 
forward by Einstein. However, it is significant for our purposes that nothing is said in Tatiana Ehrenfest's paper about Einstein's dimensional analysis of 1925.

\subsection{The adiabatic transformation and the field of conservative forces}

	Let's go back to the paper of 1925. In it, Einstein pointed out that the dimensional analysis 
with his initial assumptions has produced only his research---expression (\ref{eq:rhopsi2args})---,
but that it is possible to go further without making ``doubtful hypotheses.''%
\footnote{``ohne Setzung irgendwie zweifelhafter Hypothesen'' \cite[p.~20]{EinsteinA1925Quantentheorie2}.}
He proposed two independent 
assumptions, which both lead to the same result. He analyzed how the distribution function $\rho$ is 
modified by an adiabatic compression and by an external field of conservative forces. Referring 
to both assumptions, he wrote:

\begin{quote}
...but they are very natural, and their correctness is made more probable moreover by the fact 
that they lead to the same result and that they lead to Maxwell's distribution in the limiting case 
of vanishing quantum influence.%
\footnote{\selectlanguage{german}
	``... dieselben sind aber sehr nat\"urlich, und ihre Richtigkeit wird au{\ss}erdem
	noch dadurch wahrscheinlich gemacht, da{\ss} sie beide zu demselben Ergebnis f\"uhren,
	und da{\ss} sie in dem Grenzfalle verschwindenden \emph{Quanteneinflusses} zur
	{\scshape Maxwellschen} Verteilung f\"uhren.'' 
	\cite[p.~20]{EinsteinA1925Quantentheorie2}, our emphasis.}
\end{quote}

In order to calculate the change of (kinetic) energy during the adiabatic 
compression, Einstein resorts to the variation of the molecules' momenta in their collisions 
against the (mobile) walls. Desalvo found it ``remarkable'' that starting from such 
assumption ``Einstein obtained the correct dependence of kinetic energy on volume.''%
\footnote{\cite[p.~525]{DesalvoA1992Constant}.}
He apparently referred to the use of the relation between the change in kinetic energy and the 
change in the volume, i.e., to expression (\ref{eq:DeltaL}).
To obtain it, Einstein made an intermediate step, in which he wrote
\begin{equation} 
\Delta |p_1| = -|p_1|\frac{\Delta l_1}{l_1}.
\label{eq:Deltap}
\end{equation}
He claimed that this expression is obtained ``easily'' by applying the laws of elastic collisions. The calculation proceeds indeed straightforwardly from the consideration of energy and momentum consideration for the case of a material particle bouncing off an infinitely heavy moving piston.%
\footnote{Very similar considerations were applied by Hilbert in his use of Ehrenfest's adiabatic hypothesis for the derivation of the black-body energy density, see \cite[pp.~484--500]{SauerTEtal2009Lectures}.}
It is in this way how, invoking the laws 
of elastic collisions, a correct expression for the dependence of the kinetic energy on volume can be obtained.%
\footnote{For an elementary discussion of adiabatic compression at both the thermodynamic and
	the molecular level, see \cite{MirandaEN2002Compression}.}
However, in the previous installments of his quantum gas theory, Einstein had already deduced that the relation between pressure and energy density for the quantum gas was the same as that in the classical gas (see eq.~(\ref{eq:penerg})). Starting from this result, one can immediately calculate the variation of kinetic energy with volume, without using (\ref{eq:Deltap}). Only in this sense was the use of elastic collisions by Einstein justified in the non-statistical paper.

Picking up the comparison we proposed earlier with Ehrenfest's paper, it must be noted that 
Ehrenfest---in 1911 but also in the subsequent years in which he developed his 
idea---always referred to adiabatic transformations as pertaining to mechanical systems. Starting from 
mechanical variations he established connections between (and sometimes discovered) allowed 
quantum motions. Then, once he had a hold on the possible motions of the system, he 
calculated the most probable distribution of states among them and, postulating that this was 
the state of equilibrium, introduced the notion of temperature. The ideal gas has no mechanical 
invariants, since those can only be defined properly for periodic motion.%
\footnote{See \cite{PerezEC2009Hypothesis}.}
The adiabatic 
transformation considered by Einstein in 1925 is directly applied to a macroscopic system, i.e. to 
a thermodynamic system, in an analogous way as he proposed to do it---but did not elaborate on---in 1905 
with black-body radiation.%
\footnote{See \cite{Einstein1905i} and the discussion above.}

It would be wrong to say that Einstein made use of Ehrenfest's adiabatic hypothesis in this 
paper, since that would imply---at the very least---that one had identified connected allowed 
quantum motions. His procedure is closer to the demonstrations of Wien's displacement law in 
the late nineteenth century.

More significant is what Einstein assumed in the second path to reduce the number of 
arguments of the density function. Ignoring again collisions between molecules, he assumed 
that their motions be governed by Hamilton's equations, i.e., that they follow the laws of classical 
mechanics. In this case, the stationarity of the density function implies, as was known, its 
dependence on the Hamiltonian, i.e., on the sum of kinetic and potential energy. This result is 
generally known as Liouville's theorem. Thus, the analysis of the system in an external field 
allowed Einstein to say something about the dependence of the density on the volume. 

Note that due to the failure of the Bohr-Sommerfeld theory to account for many-electron 
systems, or to the surprising result provided by the experiment of Stern and Gerlach (in the 
analysis of which Einstein and Ehrenfest themselves had contributed%
\footnote{\cite{EinsteinAEtAl1922Bemerkungen}.}), in early 1925 the validity 
of Hamiltonian mechanics was seriously questioned. For this reason, it is surprising that Einstein 
assumed for free particles the validity of classical mechanics.
Schematically, these are Einstein's assumptions in this respect:

\bigskip
\begin{tabular}{|p{5cm}|p{5cm}|}
\hline 
	Interactions between molecules and container walls
	& Classical elastic collisions \\
\hline
	Interactions between molecules 
	& Not taken into account\\
\hline
	Free motion of the molecules 
	& According to classical mechanics\\
\hline
\end{tabular}

\bigskip

These assumptions constitute the definition of an ideal gas. They are, in other words, the 
necessary assumptions needed to obtain---in statistical mechanics---the relation between total 
kinetic energy and pressure provided by the virial theorem. Therefore, in utter contradiction to 
his original intention---and, probably, in awareness of this contradiction---, Einstein, it seems, left 
only one possibility open to right the wrong: statistics. Only in the particular way of counting 
states, in the transition from mechanics to thermodynamics, could differences between classical 
and quantum ideal gases be placed. 

However, there was another possibility: to build a new mechanics. The first assumption listed in the 
table was valid in Einstein's theory. The second assumption represents the definition of an ideal gas. Finally, the third and last assumption did not play a decisive role; without it, and using only 
dimensional analysis, very similar conclusions could be obtained. But Einstein did not make explicit in the paper any conclusion of this kind.

\section{An ignored attempt}

Let us now discuss the immediate contemporary reactions to Einstein's non-statistical paper.
First discussions around the question as to how to apply quantization to ideal gases go back to 
papers by Otto Sackur in 1911, and Hugo Tetrode in 1912.%
\footnote{\cite{SackurO1912Anwendung}, 
\cite{TetrodeH1912Konstante1,TetrodeH1912Konstante2}.}
The subject gained attention again 
in the twenties, in the course of the developments of theory and also due to the appearance of a 
widely discussed paper by Ehrenfest and Trkal.%
\footnote{\cite{EhrenfestP1920Deduction}.}
The justification of the factorial $N!$ in the 
partition function was a most widely debated issue. We are not going to analyze this episode 
here, but refer interested readers to works we have already cited.%
\footnote{For instance \cite{DarrigolO1991Statistics}, \cite{DesalvoA1992Constant}.}

Among the physicists who immediately reacted to the appearance of the series of papers by 
Bose and Einstein, Desalvo mentions Adolf Smekal and Pascual Jordan.%
\footnote{\cite[pp.~529--531]{DesalvoA1992Constant}.}
Their respective papers were received by \emph{Zeitschrift f\"ur Physik} in early July 1925.%
\footnote{\cite{JordanP1925Gleichgewicht}, \cite{SmekalA1925Beitraege}.}
We find in both papers a 
favourable disposition towards the new statistics.

On Smekal, a physicist well versed with statistical problems (and also an expert in the 
meaning and applications of adiabatic transformations), Einstein's third article did not seem to 
have left any impression. In his paper, he even affirmed that the compatibility of any statistical 
treatment with the second law of thermodynamics needs the adiabatic invariance of statistical 
weights.%
\footnote{\cite{SmekalA1926Grundlagen}.}
But neither with this point nor with others did Smekal make any explicit reference to 
Einstein's recent use of adiabatic processes. In an extensive article written for the
\emph{Encyklop\"adie der Mathematischen Wissenschaften}, 
completed shortly before, he expressed his opinion that 
the success of the Bose-Einstein approach would be decided by future experiments, i.e., that 
only future research might corroborate or reject it.%
\footnote{\cite[p.~1214]{SmekalA1926Grundlagen}.}
This would seem to be a suitable place to 
mention Einstein's non-statistical arguments in support of the theory, but there is not a single 
word about them (Smekal gives the reference to Einstein's third paper, but he does not give 
details nor even refers to any of its content).

The same is true for Jordan, whose interest was focussed on the application of Einstein's new 
results to the study of the equilibrium between matter and radiation (he also made use of the 
probability transitions introduced by Einstein \cite{Einstein1916j,Einstein1916n}). Jordan also cited the third paper but did 
not engage with its argument.

From Jordan we have available his direct testimony, obtained by T.S.~Kuhn for the \emph{Archive 
for History of Quantum Physics}.%
\footnote{Interview with P.~Jordan by T.S.~Kuhn, 18 June 1963. Microfilm transcription in 
	AHQP/OHI-3. \label{note:InterviewJordan}}
The German physicist recalled that, during the first half of 
1925, there were not many G\"ottingen physicists particularly interested in Einstein's new gas 
theory, similar to the way it had happened twenty years before with the hypothesis of light 
quanta. Nevertheless, during Ehrenfest's annual visit to G\"ottingen he provoked a debate about 
the new statistics. It is likely that in a presentation,%
\footnote{See footnote \ref{note:Ehrenfestpresentation}.}
besides explaining his own 
analysis of the fluctuations (which a little later would turn into a publication, and which Jordan 
would translate into the quantum---matrix---formalism%
\footnote{See \cite{BornM1926Quantenmechanik}, \cite{EhrenfestP1925Energieschwankungen},
\cite{DuncanAEtAl2008Resolution}}), Ehrenfest elaborated on various aspects 
of Einstein's theory, such as the loss of independence of molecules or the abrupt increase of 
concentration in the fundamental state below a certain temperature. We do not know whether 
he mentioned any of the arguments of Einstein's third non-statistical paper.

Planck and Schr\"odinger were more involved in the debates around the ideal gas and its 
quantum theory. Both authors published papers soon after the appearance of those by Einstein. 
Planck presented a communication to the Prussian Academy on 2 February 1925, less than a 
month after Einstein had presented his third paper.%
\footnote{\cite{PlanckM1925Frage}.}
Far from being a reaction to Einstein's 
theory, Planck recapitulated his previous works, in response to the experimental results by Stern 
and Gerlach. According to his opinion, the experiments with silver ions had proved that under 
certain conditions only certain paths in phase space were admissible. With this statement, he 
was renouncing the kind of quantization that he had defended in his second theory, where only 
the emission process was quantized, but not the absorption nor the mechanical motion itself.

At the end of his paper, Planck commented on Einstein's theory, pointing out its prediction 
of a loss of statistical independence of molecules. He neither criticized nor supported the new 
approach, and limited himself to the remark that the experiments will put it into its right place. 
But he also pointed out that it would imply a ``fundamental modification'' of the current ideas on 
the nature of molecules.  

Some months later, and after he had presented on July 23 to the Academy a paper by 
Schr\"odinger on the statistical entropy definition for the ideal gas,%
\footnote{\cite{SchroedingerE1925Bemerkungen}.}
Planck went back to the 
question of the entropy of an ideal gas and to Einstein's new theory.%
\footnote{\cite{PlanckM1925Entropiedefinition}.}
Here, as in his last paper 
on the subject published in \emph{Zeitschrift f\"ur Physik}, Planck only referred to the definition of 
entropy, that is, to the problem of counting and assigning probabilities.%
\footnote{\cite{PlanckM1926Definition}.}
But in these papers, with 
which Planck closed a long series of works dedicated to the study of the ideal gas under the 
new light of the quantum theory, there is not a single reference to the theoretical tests proposed 
by Einstein in his third paper. 

Something similar happened in the case of Schr\"odinger.%
\footnote{About Schr\"odinger's work on quantum statistics of ideal gases (and also about Planck's),
	see \cite{HanlePA1977Coming}.}
Recall that it had been Einstein's 
theory, and the ensuing epistolary exchange between both physicists, which had turned 
Schr\"odinger's attention to de Broglie's work. The most significant paper of Schr\"odinger, for our 
present concern, is ``on Einstein's gas theory.''%
\footnote{\cite{SchroedingerE1926Gastheorie}.}
In it, he arrived at the same results achieved by 
his colleague, but he applied Boltzmann's statistics to waves \`a la de Broglie, i.e., he treated 
particles as excitations. In the interesting paper that Planck presented on his behalf to the 
Prussian Academy on July 23, Schr\"odinger recognized that the compatibility of Bose-Einstein's 
distribution with Nernst's theorem was an important point in favour of the new theory.%
\footnote{\cite{SchroedingerE1925Bemerkungen}.}
In this 
paper, Schr\"odinger analyzed different entropy functions which had been used in the past or 
were being used at the moment.

Einstein himself argued in favour of his theory appealing to the third paper in a letter, with which 
he answered some of Schr\"odinger's criticism. Schr\"odinger's letter, on the other hand, showed that 
he had not understood the new way of counting which was implicit in Bose's statistics.%
\footnote{See Schr\"odinger to Einstein, 5 February 1925 (AEA~22-001).}
In his response Einstein had to point out:

\begin{quote}
In Bose's statistics, which I use, the quanta or molecules are regarded as \emph{not independent of each other}. [...] I failed to emphasize clearly the fact that here a new kind of statistics is employed, which for the time being is justified by nothing but its success.%
\footnote{\selectlanguage{german}
``In der von mir verwendeten Bose'schen Statistik werden die Quanten bzw.\ Molek\"ule nicht
als \emph{voneinander unabh\"angig} behandelt. [...] Ich verabs\"aumte es, deutlich hervorzuheben, dass hier eine besondere Statistik angewendet ist, die durch nichts anderes als durch den Erfolg
vorl\"aufig begr\"undet werden kann.'' Einstein to Schr\"odinger,
	28 February 1925 (AEA~22-002). For a French translation of this letter, see
	\cite[p.~194]{OCAE01}.}
	\end{quote}
And about the third, non-combinatorial paper, he wrote:

\begin{quote}
In a third paper, which is currently in press, I lay out considerations that are independent of 
statistics and that are analogous to the derivation of Wien's displacement law. These latter 
results have convinced me completely of the correctness of the road to follow.%
\footnote{\selectlanguage{german}
	``In einer dritten Arbeit, die gegenw\"artig im Druck ist, werden Betrachtungen
	gegeben, die von der Statistik unabh\"angig sind und der Abl.\ des Wienschen 
	Verschiebungsgesetzes analog sind. Diese letzten Ergebnisse haben mich von der 
	Richtigkeit des eingeschlagenen Weges fest \"uberzeugt.'' Ibid.}
\end{quote}

Once again, Einstein tried to underline the independence of statistics of the arguments in his 
paper. We will see in the following section that this was also the way he presented them to 
Ehrenfest.

As far as we have been able to determine, none of the major journals contain papers that 
called attention to the last instalment of Einstein's new theory of the quantum ideal gas, not 
even in \emph{Zeitschrift f\"ur Physik}, where many of the most relevant papers of theoretical physics of 
those days were published. We think one should bear in mind Jordan's verdict about those 
days: The statistical treatment of the ideal gas was not anything that worried the physicists.%
\footnote{See footnote \ref{note:InterviewJordan}.}
A superficial consultation of other journals seems to confirm this statement. 

Neither was there a presentation especially devoted to quantum statistics during the fifth Solvay 
conference in 1927 (Fermi-Dirac statistics had been born the previous year).%
\footnote{\cite{NN1928Electrons}, for an account of the 1927 conference, see \cite{BacciagaluppiGEtAl2009Crossraods}.}
In fact, as we will 
discuss below, it was Einstein who had been asked by Lorentz to give a talk on that subject but 
he declined the invitation a few months before.
Einstein's third paper on quantum ideal gas theory did not have any effect. Nor did the papers 
that preceded it and that even today constitute the most known and celebrated part of his theory 
cause an avalanche of reactions.

\section{Einstein, Ehrenfest, and the Bose-Einstein statistics}

Let us now look at Ehrenfest's role for Einstein's non-combinatorial paper.
There are different reasons that make it worthwhile to investigate it.
The reference to 
Ehrenfest in the second paper indicates that both colleagues had discussed the new theory 
after the publication of its first instalment, in July 1924.%
\footnote{See note \ref{note:PEmention}.}
This reference is not surprising, 
because in 1924 Einstein and Ehrenfest had a very close relation, both professionally and 
personally; in fact, since 1920 Einstein had been an official visitor staying in Leiden, coming 
from time to time to Leiden as a teacher.%
\footnote{See \cite{KleinM1985Ehrenfest,DelftD2006Einstein} and the Introduction to \cite[pp.~xlii-xlvi]{CPAE10}.}
Moreover, many of the questions Einstein's theory put 
on the table had been and still were subject of interest to Ehrenfest.

As we have already said, Ehrenfest had penetrated into the analysis of the implicit statistics in 
Planck's radiation law, comparing it with that underlying in Wien's.%
\footnote{See section 5 above.}.
Iuri A.~Krutkow, one of Ehrenfest's students from his Petersburg 
years, had elucidated this 
contraposition, during one of his stays in Leiden together with Ehrenfest, even more in a polemic he maintained in \emph{Physikalische Zeitschrift} with the Polish 
physicist Mieczys{\l}aw Wolfke.%
\footnote{\cite{KrutkowI1914Annahme}, \cite{KrutkowI1914Bemerkung}.}
The contraposition presented by Einstein in 1925 between Boltzmann's and Bose's 
statistics strongly recalls these papers of the Russian physicist.

In general, it was Ehrenfest who had extracted more information from Wien's displacement law 
in his quantum researches than anyone else. It has been observed that it was precisely 
the analysis of Wien's law that had led him to the adiabatic hypothesis.%
\footnote{\cite{KleinM1985Ehrenfest}, \cite{NavarroLEtal2004Ehrenfest,NavarroLEtal2006Ehrenfest}.}
In Einstein's third paper the main character is Wien's displacement law. 

In addition, in 1905 Ehrenfest maintained the polemic with Jeans on the dimensional 
derivation of the displacement law we have analyzed in Section 5.2, which inspired Einstein's own derivation  in 1909. 
Hanle suggests, commenting on the statistical works of Schr\"odinger prior to the formulation of 
wave mechanics, that what we are calling here Einstein's third paper might be interpreted as a 
response specifically designed to convince Ehrenfest of the suitability of the new theory.  
Although Hanle maintains that Ehrenfest's attitude should be taken as representative of that of 
other physicists, the third paper would be constituted, according to this interpretation, by a new 
series of arguments put together in order to convince Ehrenfest of the plausibility of the 
obtained results. This opinion is supported by documentary evidence in some letters.%
\footnote{See footnote \ref{note:inkri}.}

Not any less important is the talk Ehrenfest gave in G\"ottingen at the beginning of the summer of 
1925. Among other things, he spoke about the new statistics of Bose and Einstein. According to 
Jordan, Ehrenfest showed himself to be sceptic of the new method.\footnote{See footnte \ref{note:InterviewJordan}.}
His presence---presumably during June---left its mark in more than one physicist.
Max Born referred to it in a letter to Einstein: 

\begin{quote}
... your brain, heaven knows, looks much neater: its products are clear, simple, and to the 
point. With luck, we may come to understand them in a few years' time. This is what happened 
in the case of your and Bose's gas degeneracy statistics. Fortunately, Ehrenfest turned up here 
and cast some light on it. Then I read Louis de Broglie's paper, and gradually saw what they 
were up to. I now believe that the wave theory of matter could be of very importance.%
\footnote{\selectlanguage{german}
	``Dein Gehirn sieht, weiss der Himmel, reinlicher aus. Seine Produkte sind klar, 
	einfach und treffen die Sache. Wir kapieren es dann zur Not ein paar Jahre sp\"ater.
	So ist es uns auch mit Deiner Gasentartung gegangen. Gl\"ucklicherweise erschien
	Ehrenfest hier und hat uns ein Licht aufgesteckt. Darauf habe ich die Arbeit von Louis
	de Broglie gelesen und bin allm\"ahlich auch hinter Deine Schliche gekommen. Jetzt
	glaube ich, dass die `Wellentheorie der Materie' eine sehr gewichtige Sache werden kann.''
	Born to Einstein, 15 July 1925 (AEA~8-177). English translation in
	\cite[p.~83]{BornM2005Letters}.}
\end{quote}
In a note in a well-known paper of 1926, in which Born together with Jordan and Heisenberg 
developed the matrix mechanics formulation of quantum theory, the authors also refer to this 
talk: ``P. Ehrenfest, Lecture in the G\"ottingen seminar on the Structure of Matter, Summer 1925. 
The contents of this lecture were of great assistance to our present considerations.''%
\footnote{\cite{BornM1926Quantenmechanik}. English translation in \cite{WaerdenBL1967Sources}; the quote
	is on p.~380, note~2.\label{note:Ehrenfestpresentation}}
According to the context in which this footnote appears, we can assume that Ehrenfest presented the 
fluctuation analysis he published a few months later.%
\footnote{\cite{EhrenfestP1925Energieschwankungen}.}

Nevertheless, we can see in their correspondence that both Einstein and Ehrenfest also rejected the 
proposal by Bohr, Kramers and Slater. On 9 January 1925, Ehrenfest wrote to his friend:

\begin{quote}
If Bothe-Geiger should find ``statistical independence'' of electrons and scattered light quantum, 
it would prove nothing. But if they find dependence, then it is a triumph of Einstein over Bohr.- 
This time I believe (but only this time!) firmly in you, that is I would be glad if dependence would 
be made evident.%
\footnote{``Falls Bothe-Geiger \glqq statistische Unabh\"angigkeit\grqq~von Elektron und gestreutem
	Lichtquant beweist es \emph{nichts}. Falls sie aber \emph{Abh\"angigkeit} finden ist es ein 
	Triumph von Einstein \"uber Bohr.--- Diesmal glaube ich (ausnahmsweise) fest an Dich,
	w\"urde mich also freuen Abh\"angigkeit evident gemacht w\"urde.'' Ehrenfest to Einstein, 
	9 January 1925 (AEA~ 10-100).}
\end{quote}

Two days later, Einstein responded to a previous letter by Ehrenfest, commenting on the fluctuation analysis of a vibrating string with which Ehrenfest had 
discussed the dual nature of the radiation suggested first by Einstein in 1909.%
\footnote{\cite{EhrenfestP1925Energieschwankungen}.}: 

\begin{quote}
I forgot in my letter to express my agreement with the statistical consideration about the 
fluctuations of the energy of the subvolume according to the understanding of standing waves. It 
would probably be good to publish this at some point.%
\footnote{\selectlanguage{german}
	``Ich verga{\ss}, Dir in meinem Briefe zuzustimmen zu der statistischen Betrachtung
	\"uber die Schwankungen der Energie des Teilraumes nach der Auffassung der stehenden 
	Schwingungen. Es w\"are wohl gut, dies einmal zu publizieren.'' Einstein to Ehrenfest,
	11 January 1925 (AEA~10-102).}
\end{quote}
Ehrenfest followed his friend's advice and sent it for publication in August. As mentioned above, its content must have been the core of the talk he gave in G\"ottingen that summer.

In this paper, Ehrenfest calculated, in different ways, the energy fluctuations of the simple 
system of a string held fixed in its two ends. In all cases, he used the normal modes with which he 
had analyzed the black-body radiating cavity more than ten years before. Thus, he showed and 
demonstrated a conclusion to which Leonard Ornstein and Frits Zernike had already come 
earlier: Einstein had supposed that the entropy was an extensive quantity, which is incompatible 
with a pure wave treatment (because of  overlapping of the waves). We will not analyze 
Ehrenfest's paper here (he already referred to ``Bose-Einstein's statistics'' and also alluded to 
the concept---but not the name---of a phonon), but we will refer to recent works that have put it in 
direct relation to a later analysis by Jordan in which many historians place the origin of quantum 
electrodynamics.%
\footnote{\cite{DuncanAEtAl2008Resolution}.}
The contradiction demonstrated by Ehrenfest turned, in the hands of Jordan 
and his quantum mechanical approach, into one of the first demonstrations of complementarity. 
Both terms in the fluctuation expression, which Einstein had attributed in 1909 to corpuscular 
and undulatory components respectively, were shown to be necessary consequences of the 
new mechanics.

In Ehrenfest's paper we see what remained of the old project of publishing a note with Krutkow 
and Bursian. It is only a footnote:

\begin{quote}
The words of the paper by S.N.~Bose, Planck's Law and the Light Quantum Hypothesis [ref.], 
readily create the impression as though Planck's radiation law could be derived from the 
assumption of \emph{independent} light corpuscles. But this is not the case. Independent light 
corpuscles would correspond to Wien's radiation law.%
\footnote{``Der Text der Arbeit von S.N.~Bose, Plancks Gesetz und Lichtquantenhypothese [ref.]
	erweckt leicht den Eindruck, als ob sich das Plancksche Strahlungsgesetz aus der Vorstellung
	\emph{unabh\"angiger} Lichtkorpuskeln ableiten lie{\ss}e. Aber das ist nicht der Fall.
	Unabh\"angige Lichtkorpuskeln w\"urden dem Strahlungsgesetz von W.~Wien entsprechen.''
	\cite[p.~364, note~1]{EhrenfestP1925Energieschwankungen}. Ehrenfest's emphasis.}
\end{quote}

It seems that Ehrenfest's influence on Einstein's research during those days was not very 
significant. From the available evidence we cannot exclude the possibility of additional meetings 
before the presentation of the second paper, but also in this case it would not appear that there 
existed a very close collaboration between the two friends. To confirm this impression, we have 
consulted the Paul Ehrenfest Archives, particularly his correspondence and notebooks.

We have found no evidence of Ehrenfest's concern with the problem of the quantum ideal 
gas in those months of 1924 and 1925. In his notebooks there are scarcely any annotations on 
this subject. There are many notes on what appears to be the calculations for his paper of 1925 
on fluctuations.%
\footnote{See ENB:1--28 and ENB:1--29, from April 1923 to December 1926. In EHA, microfilms
AHQP/EHR-4 and AHQP/EHR-5.}
These series of annotations date back to mid-December 1924. Among them we 
find references to ``Einstein's fluctuations'' and to the comments of Tatiana to which he referred 
in the paper.

	In the lists of points that indicate something like topics to be treated next we can find 
references to light quanta, e.g. in this one:

\begin{quote}
...\\
$h\nu$-corpuscles\\
$\delta Q/T$ in Quantumth.[eory]\\
...\\
Radiat.[ion]  Fluct.[uations]\\
$1/N !  \leftrightarrow$  Gibbs Paradox\\
$\rightarrow$ Bose\\
...%
\footnote{ENB:1-29, probably around 24 December. In EHA, microfilm AHQP/EHR-5.\label{note:Ehrenfestdiaries}},
\end{quote}

In any case, there is not an alluvium of annotations related to the new way of counting 
introduced by Bose. Certainly, there are isolated annotations on Bothe's works or on Gibbs's 
paradox and the $N!$-factor, but nothing more. 

Let us recall again that the statistical observations that Einstein did in his second paper on gas 
theory did not need any extensive of sophisticated calculations, at least for Ehrenfest, who was deeply convinced that only 
some kind of dependence among quanta might lead to Planck's law. We would not be surprised 
if the way, in which Einstein demonstrated the peculiarities of the new statistics was suggested to him by 
Ehrenfest. But we have no evidence for this possibility.

On the contrary, nothing seems to indicate that Ehrenfest collaborated closely with Einstein in the 
months between the arrival of Bose's letter and the publication of the third paper.%
\footnote{In a letter he wrote in May 1927, Ehrenfest asked Einstein for offprints of subsequent 
	papers to the second one of the theory of the quantum ideal gas. The non-statistical
	paper is then the first one he had no offprint of. However, we do not think this
	coincidence is significant. Ehrenfest to Einstein, 16 May 1927 (AEA~10-164).}
According to him, in those days:

\begin{quote}
About my scientific things, all goes so incredibly bad, that I would be very happy if I could 
already retire!%
\footnote{Ehrenfest to Joff\'e, 16 February 1925. In \cite[p.~186]{MoskovchenkoNI1990Ehrenfest}.}
\end{quote}

(Ehrenfest was 46 years old). The density of annotations in his 
notebooks is indeed not very high; but then annotations could have been done in  
notebooks that no longer exist, and Ehrenfest's tendency to underestimate both his capacities and 
his achievements is notorious.

\section{A rightfully forgotten paper?}

This question has two opposite answers, depending on whether one refers it to the 
reception that Einstein's work had in the months immediately after its publication, or to the 
attention it has received in later years by historians of quantum physics.

In the first case the oblivion seems understandable. The practically immediate appearance 
of the revolutionary contributions of 1925 to quantum theory eclipsed any possible interest of 
Einstein's paper. The arguments it contains only concern the ideal gas from a thermodynamic 
perspective. But, what is more important, it includes hypotheses that were in open contradiction 
with the course quantum researches had taken. Many physicists had rejected already the laws 
of mechanics, and Einstein assumed their validity for describing the motions of the gas 
molecules.

The papers of the twenties that refer to Einstein's theory usually 
mention all three instalments. This indicates that, in spite of the almost complete lack of 
comments on it, its existence was known. We are inclined to think that it simply was not of any 
interest to Einstein's colleagues. Einstein justified the considerations of the non-statistical paper 
with the deep dissatisfaction over the statistical route by which he had arrived at the new 
distribution function. However, the problem was not whether his colleagues saw Bose's statistics 
favourably, but that in the following months the physicists' ideas around the quantum issues 
changed substantially. Bose's statistics, in spite of implying a way of counting that 
was incompatible with classical statistics, led to an already accepted result. This was much 
more than could be said of other attempts of explaining, for example, the Zeeman effect or the multielectronic 
spectra. This state of affairs appears to be what Niels Bohr was referring to in a postscript he added to a paper, after 
learning about the result of the Bothe-Geiger experiments, in July 1925:

\begin{quote}
The renunciation of space-time pictures is characteristic of the formal treatment of problems of 
the radiation theory and of the mechanical theory of heat attempted in recent papers by de 
Broglie and Einstein. Especially in consideration of the perspectives opened up by these papers, 
I have thought that the discussion presented in the preceding paper might be of some interest, 
and I have therefore decided to publish the paper without change, although the endeavour 
underlying it may now seem hopeless.%
\footnote{\cite{BohrN1925Wirkung}. English translation in \cite[p.~206]{StolzenburgK1984Bohr}.}
\end{quote}

Although the Bothe-Geiger results supported the light quanta hypothesis before the BKS theory, 
Bohr insisted on the necessity of giving up the ``space-time'' pictures. As an example, he gave 
precisely Einstein's theory of the quantum ideal gas. But Bohr's attitude towards Einstein's quantum theory was biased. He mentioned de Broglie's 
dissertation and only the first two papers of Einstein's theory. This omission might be deliberate, 
since the third paper does not fit in that ``renunciation'' that Bohr alluded to. Einstein's paper 
begins as follows:

\begin{quote}
Motivated by a derivation of {\scshape Planck}'s radiation formula, which was given by {\scshape Bose}
and which is based stringently on the hypothesis of light quanta, I recently formulated a
theory of the ideal gas. This theory appears justified, if one proceeds on the assumption that 
a light quantum differs (apart from its property of polarization) from a monatomic molecule
essentially only by the vanishingly small rest mass of the quantum.%
\footnote{\selectlanguage{german}
	``Angeregt durch eine von {\scshape Bose} herr\"uhrende Ableitung der 
	{\scshape Planck}schen Strahlungsformel, 
welche sich konsequent auf die Lichtquantenhypothese st\"utzt, habe ich neulich eine 
Quantentheorie des idealen Gase aufgestellt. Diese Theorie erscheint dann als berechtigt, 
wenn man von der \"Uberzeugung ausgeht, da{\ss} ein Lichtquant (abgesehen von seiner 
Polarisationseigenschaft) sich von einem einatomigen Molek\"ul im wesentlichen nur dadurch 
unterschiede, da{\ss} die Ruhemasse des Quants verschwindend klein ist.''
\cite[p.~18]{EinsteinA1925Quantentheorie2}.}
\end{quote}

As we have seen, Einstein also assumed the validity of Hamilton's equations for the mechanical 
motion of the gas molecules. Therefore, Einstein took as a starting point what the Danish 
physicist propagated to ``renunciate.''

In retrospect, Einstein's initial suspicion about Bose's statistics will turn into one of the first 
symptoms of his later distancing himself from quantum mechanics. For this reason we find no 
justification for the neglect of Einstein's paper by historians of physics. Perhaps we are dealing 
here with Einstein's last attempt to contribute positively to the construction of the quantum 
theory, for which he had done so much. In addition, this paper closed the circle he initiated in 
1905 with the hypothesis of energy quanta. First, the analogy was going one way, now, finally, it 
was also going the other way. The statistical dependence among light quanta which had limited 
the analogy with an ideal gas now was found also among molecules. Hence, for the first time 
the analogy was complete.

In the months before the fifth Solvay conference, which was devoted to photons and electrons, 
Einstein declined Lorenz's invitation to give a talk on quantum statistics. This happened a bit 
more than two years after his trilogy on the ideal quantum gases was published. These are 
Einstein's words:

\begin{quote}
I recall having committed myself to you to give a report on quantum statistics at the Solvay 
congress. After much reflection back and forth, I come to the conviction that I am not competent 
[to give] such a report in a way that really corresponds to the state of things. The reason is that I 
have not been able to participate as intensively in the modern developments of the quantum 
theory as would be necessary for this purpose. This is in part because I have on the whole too 
little receptive talent for fully following the stormy developments, in part also because I do not 
approve of the purely statistical way of thinking on which the new theories are founded.% 
\footnote{\selectlanguage{german}
	``Ich erinnere mich, dass ich Ihnen gegen\"uber die Verpflichtung \"ubernommen
	habe, am solvay-Kongress ein Referat zu halten \"uber Quanten-Statistik. Nach vielem
	Hin- und Her-\"Uberlegen komme ich aber zu der \"Uberzeugung, dass ich nicht f\"ahig 
	bin zu einem solchen Referat, das wirklich dem Stande der Dinge entspricht.
	Der Grund liegt darin, dass ich die moderne Entwicklung der Quantentheorie nicht so
	intensiv habe mitmachen k\"onnen, wie es hiezu n\"otig w\"are. Das kommt teilweise daher,
	dass ich \"uberhaupt receptiv zu wenig begabt bin, um der st\"urmischen Entwicklung
	v\"ollig zu folgen, teilweise auch daher, weil ich innerlich die rein statistische
	Denkweise, auf denen die neuen Theorien beruhen, nicht billige.''
	Einstein to Lorentz, 17 June 1927 (AEA~71-153). In 
	\cite[pp.~431--432]{PaisA1982Subtle}.}
\end{quote}

In that last year Born had proposed the probabilistic interpretation of the wave function. Surely 
Einstein was thinking of that result, but arguably he must also bear in mind that the only possible 
characterization of the quantum ideal gas was still statistical. 

Ehrenfest shared this rejection. He had noticed several years ago what now appeared 
clearly in Einstein's theory: If the particles were treated as statistically independent ones the law 
observed and confirmed in the laboratory could never be derived. But if anything appears with 
clarity after examining the correspondence between the two friends, and also their publications, 
it is is that not for a glimpse they conceived anything similar to the \emph{indistinguishability} of the 
particles. That is to say, they always thought in terms of a certain ``statistical dependence.'' What 
did not even cross their minds was to think that the way of counting introduced by Bose---and, in 
a certain way, already by Planck---could be, in fact, a new way of counting.

This is more noticeable if we remember the thesis argued by Don Howard.%
\footnote{\cite{HowardD1990Prehistory}.}
According to 
him, Einstein was also aware, since 1909, that statistically independent quanta would never lead 
to Planck's radiation law. Until he received Bose's letter and manuscript, he had never applied the analogy from 
quanta to molecules. Therefore, indistinguishability was far from being born in 1924. 
Furthermore, Howard claims that it was entanglement which mainly worried Einstein in the new 
mechanics, not probability. Expressions used by Einstein, like ``incriminated statistics'' or ``purely 
statistical way of thinking''  as well as the goal of the non-statistical paper we 
have analyzed support this claim.

However, Einstein became more and more convinced of the good sense of his approach to 
the quantum ideal gas. He was interested in new experimental results to test the quantum 
corrections.%
\footnote{See Einstein to Kamerlingh Onnes, 4 November 1924; Kamerlingh Onnes to Einstein,
	13 November 1924 (AEA 14-384, 14-386).}
Only the statistical side deserved contemptuous comments by his author. As he 
told to Ehrenfest, he tried to ensure the macroscopic (thermodynamic) facts in order to build---or 
to have an intuition of---the mechanics behind them. Only a new mechanics would square with 
an ideal gas with such odd properties.

But Einstein's paper also displayed an ambiguity, which also would contribute to his colleagues' lack 
of interest in his third paper on the subject. The ambiguity comes to the fore in his questioning 
the very concept of ideal gas, and this questioning was precisely the idea under analysis. For 
instance, in the introduction to the paper, Einstein announced he would not assume that collisions 
between molecules are governed by mechanics. In fact, what he did in the paper is, to neglect them, as corresponds to an ideal gas. But neither Einstein nor Ehrenfest appear to have given up the expectation that Bose's counting could conceal some kind of `quantum influence,' which prevented 
one from talking properly of an ideal gas; Einstein considered some kind of ``thermal forces.''%
\footnote{See note \ref{note:Grommer}.}
The waves, and particularly the way in which de Broglie introduced them in the quantum 
treatments, might allow a recompositon of the puzzle into which the ideal quantum gas had 
turned. Einstein's initial preference for Schr\"odinger's wave mechanics, in opposition to the 
matrix mechanics of Heisenberg, Born and Jordan, is as much known as understandable.

The last ``positive contribution'' of Einstein to 
statistical physics includes a paper in which he offered arguments independent of the 
``incriminated statistics,'' because what nowadays is called Bose-Einstein's statistics was not 
more, according to its creator, than a calculatory artifice absolutely devoid of any physical 
meaning. It was simply a consequence of using the wrong mechanics or of not 
considering some kind of interaction. As Einstein explained to Halpern, it ``cannot be considered as 
giving a true theoretical basis to Planck's law.'' 

\bigskip

\emph{Acknowledgments:} We thank Luis Navarro for reading an early version of the manuscript and making  
interesting comments on it. The focus of our attention on the third paper by Einstein on the 
quantum ideal gas, is due to his suggestion. Conversations 
with Pere Seglar have also helped  very much in elucidating some statistical (and
non-statistical) points discussed in this paper. T.S.\ thanks the Max Planck Insitute for the History of Science and its history of quantum physics group for their hospitality in the summer 2009.

\bigskip

The following abbreviations have been used:
\begin{tabbing}
xxxxxx\= \kill\\
AEA\>	Albert Einstein Archives, The Hebrew University of Jerusalem, Israel. \\
	\> Unpublished correspondence quoted by permission.\\
AHQP\>	Archive for History of Quantum Physics. \\ 
	\>For a catalogue, see \cite{KuhnTEtAl1967Sources}.\\
EHA\>	Ehrenfest Archive, Rijksarchief voor de Geschiedenis van de\\
	\>Natuurwetenschappen en van Geneeskunde, Leiden, Netherlands.\\
	\>For a catalogue, see \cite{WheatnB1977Catalogue}.\\
	\>We quote from the microfilm version included in the AHQP.\\
HPE\>   Huisbibliotheek van Paul Ehrenfest, Institut Lorentz, Leiden,\\
	\>Netherlands.
\end{tabbing}

%\nocite{*}   % this will print the entire bibliography

\bibliographystyle{kluwer}
\bibliography{refs}

\end{document}